\documentclass[aps,pra,floatfix,twocolumn,superscriptaddress,showpacs]{revtex4}
\pdfoutput=1
\usepackage{graphicx}
\usepackage{amssymb}
\usepackage{amsmath}

\newcommand {\ket}[1] {|#1 \rangle}
\newcommand {\bra}[1] {\langle#1 |}

\newcommand {\av}[1] {\langle #1 \rangle}
\newcommand{\rs}{\rm \scriptscriptstyle}

\begin{document}

\title{Physical replicas and the Bose-glass in cold atomic gases}

\author{S.~Morrison}
\author{A.~Kantian}
\affiliation{Institute for Theoretical Physics, University of
Innsbruck, Technikerstr. 25, A-6020 Innsbruck, Austria, \\ and
Institute for Quantum Optics and Quantum Information of the Austrian
Academy of Sciences,  A-6020 Innsbruck, Austria}
\author{A.~J.~Daley}
\affiliation{Institute for Theoretical Physics, University of
Innsbruck, Technikerstr. 25, A-6020 Innsbruck, Austria, \\ and
Institute for Quantum Optics and Quantum Information of the Austrian
Academy of Sciences,  A-6020 Innsbruck, Austria}

\author{H.~G.~Katzgraber}
\affiliation{Theoretische Physik, ETH Zurich, CH-8093 Z\"urich,
Switzerland}
\author{M.~Lewenstein}
\affiliation{ICFO-Institut de Ci\`encies Fot\`oniques, Parc Mediterrani
de la Tecnologia, E-08860 Castelldefels (Barcelona), Spain, and \\
ICREA - Instituci\`o Catala de Ricerca i Estudis Avan{\c c}ats,
08010 Barcelona, Spain}

\author{H.P.~B\"uchler}
\affiliation{Institute for Theoretical Physics III, University of
Stuttgart, Pfaffenwaldring 57, 70550 Stuttgart, Germany}

\author{P.~Zoller}
\affiliation{Institute for Theoretical Physics, University of
Innsbruck, Technikerstr. 25, A-6020 Innsbruck, Austria, \\ and
Institute for Quantum Optics and Quantum Information of the Austrian
Academy of Sciences,  A-6020 Innsbruck, Austria}

\date{\today}

\begin{abstract}

We study cold atomic gases in a disorder potential and analyze the
correlations between different systems subjected to the same disorder
landscape. Such independent copies with the same disorder landscape
are known as replicas. While in general these are not accessible
experimentally in condensed matter systems, they can be realized using
standard tools for controlling cold atomic gases in an optical lattice.
Of special interest is the overlap function which represents a natural
order parameter for disordered systems and is a correlation function
between the atoms of two independent replicas with the same disorder.
We demonstrate an efficient measurement scheme for the determination
of this disorder-induced correlation function.  As an application, we
focus on the disordered Bose-Hubbard model and determine the overlap
function within perturbation theory and a numerical analysis. We
find that the measurement of the overlap function allows for the
identification of the Bose-glass phase in certain parameter regimes.

\end{abstract}

\pacs{03.75 Lm, 64.60 Cn, 42.50 -p}
\maketitle

\section{Introduction}
\label{sec:intro}

The interplay between disorder and interaction gives
rise to a plethora of fundamental phenomena in condensed
matter systems. The most predominant examples include
spin glasses \cite{binder:86,mezard:87,young:98}, the
superconducting-to-insulator transition in thin superconducting
films \cite{finkelstein:94,fisher:90}, and localization
phenomena in fermionic systems such as weak localization and the
metal-insulator transition \cite{lee:85}. While the nature of order
in spin glasses and its theoretical description is still highly debated
\cite{parisi:79,mezard:84,bray:86,fisher:86,binder:86,huse:87,newman:96,krzakala:00,palassini:00},
a substantial contribution to the understanding of bosons
in a disordered medium has been provided by work on the
disordered Bose-Hubbard model introduced by Fisher {\em et
al.}~\cite{fisher:89}. The disordered Bose-Hubbard model has
recently attracted considerable attention due to its potential
realization with cold atomic gases in an optical lattice
\cite{damski:03,sanpera:04,roth:03,roth:03a,bloch:07,lewenstein:07,jaksch:04}.

\begin{figure}[t]
\centerline{\includegraphics[width=6cm]{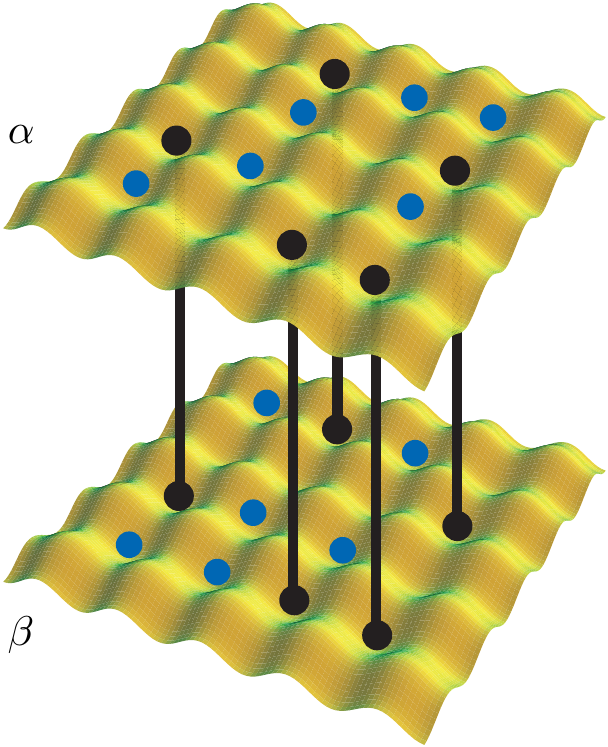}}
\caption{(Colour online)
Schematic representation of the implementation of disorder replicas
with the cold atomic gases toolbox: Two planes $\alpha$ and $\beta$
with equal disorder realisations, illustrated here for the case
of a disorder landscape introduced using a second particle species
(black dots), in which additional probe atoms (blue/lighter dots)
evolve. The planes can be combined to measure correlation functions,
such as the Edwards-Anderson overlap function.
} \label{Fig:twolayers}
\end{figure}

A general challenge in the description of glass phases in disordered
media is the absence of a simple order parameter distinguishing
the different ground states. This problem becomes evident in the
disordered Bose-Hubbard model where the phase diagram is determined
by the competition between a superfluid phase, the Mott-insulator,
and a Bose-glass phase. While the superfluid phase exhibits a
finite condensate fraction and is characterized by off-diagonal
long-range order, the Bose glass is only distinguished from the
incompressible Mott-insulator by a vanishing excitation gap and a
finite compressibility. However, in experiments on cold atomic gases
the excitation gap and the compressibility are difficult to determine
accurately and are also obscured by the finite harmonic trapping
potential. The challenge is therefore to develop measurement schemes
allowing for the experimental determination of these observables
or to develop new observables to characterize the glass phase.
While inaccessible in ``real materials,'' the Edwards-Anderson
order parameter \cite{edwards:75,binder:86} is commonly studied
analytically and measured numerically to quantify the ``order'' in a
disordered system.  It can be expressed as the correlation between
independently evolving systems with the same disorder landscape
(so-called replicas), or as a correlation of the same system at
different temporal measurements. Because the latter depends on an
extra variable (temporal measurement window) it is advantageous
to study two replicas of the system with the same disorder. Thus,
the measurement of this order parameter requires the preparation of
several samples with exactly the same disorder landscape, and the
subsequent measurement of correlations between these.

We demonstrate that such a procedure is feasible in cold atomic
gases allowing one to gain access to characteristic properties of
disordered systems naturally hidden in condensed matter systems. The
basic idea is to focus on cold atomic gases in an optical lattice:
along one direction, the optical lattice is very strong and
divides the system into independent two-dimensional (2D) planes
\cite{stoeferle:04,sebby-strabley:06,foelling:07,spielman:07}. In
addition, the system is subjected to a disorder potential and we are
interested in the situation where in each plane the same disorder
landscape is realized (see figure~\ref{Fig:twolayers}). Although the
atoms in different planes are decoupled, the presence of the same
disorder landscape within each plane induces a correlation between the
different realizations: this correlation is of special interest in the
replica theory of spin glasses and allows one to measure the overlap
function, a characteristic property of spin glasses.  Furthermore,
we show that this correlation function is accessible in experiments
on cold gases:  The main idea is to quench the motion of the atoms
by a strong optical lattice, and combine the different planes into
a single one. Subsequently, the particle number occupation within
each well carries the information about the correlation function.
The possibility for the accurate determination of the particle number
within each well of an optical lattice has recently been demonstrated
experimentally for the superfluid-Mott-insulator quantum phase
transition \cite{campbell:06a}.

Note that this measurement scheme for the overlap between system
with the same disorder can be applied to any disordered one- or
two-dimensional system realized with cold atomic gases in an optical
lattice. Of special interest is the realization of spin glasses and
their study in the quantum regime.  As an application, we focus here on
the disordered Bose-Hubbard model and calculate the overlap function
analytically in different regimes and compare it with one-dimensional
numerical simulations. We find that the overlap function $q$ exhibits
a sharp crossover from the Mott-insulating phase with $q\approx 0$ \cite{comment:overlap} to
the Bose-glass phase with $q \approx p(1-p)$, with $p\in(0,1)$, and thus
makes it possible to distinguish the two phases. Because the superfluid
phase can be detected via the interference peaks in a time-of-flight
measurement, we propose that this novel measurement scheme for the
overlap function can be used for a qualitative experimental verification
of the phase diagram for the disordered Bose-Hubbard model.

Different implementations of disorder or quasi-disorder
in cold atomic gases have been discussed and experimentally
realized. Several groups attempted to search for traces of Anderson
localization and the interplay of disorder-nonlinear interactions
in Bose-Einstein condensates (BECs). The first experiments
\cite{lye:05,fort:05,clement:05,clement:06,schulte:05} were
performed with laser speckles that had a disorder correlation
length larger than the condensate healing length. Localisation
effects were thus washed out by interactions. As an alternative,
superlattice techniques---which are combinations of several optical
potentials with incommensurable spatial periods---were used to
produce quasi-disordered potentials with short correlations. This
approach allowed the observation of some signatures of the Bose
glass by the Florence group \cite{fallani:07}. Only very recently
the Palaiseau group has managed to create speckle potentials on the
sub-micron scale and directly observe Anderson localisations effects
in a BEC released in a one-dimensional waveguide \cite{billy:08} based on
the theoretical predictions of \cite{sanchez:07,sanchez:08}. The
Florence group reported observations of localisations phenomena in
quasi-disordered potentials in BEC of K$^{39}$, which allows for
complete control of the strength atomic interactions using Feshbach
resonances \cite{roati:08}. It is worth noting that experimental
attempts for local addressability in an optical lattice also offer the
possibility to create disorder with correlation lengths comparable to
the lattice spacing \cite{lee:07,gorshkov:07}. Mixtures of cold gases,
on the other hand, provide an alternative approach for the realization
of disorder by quenching the motion of one species by a strong optical
lattice, which then provides local impurities for the second species
\cite{vignolo:03,gavish:05}. While production of 2D planes with equal disorder
realisations follows naturally if the disorder is implemented with
laser speckles, we also show how such a system can be realized in
the case of disorder induced by a mixture of atomic gases.

In section~\ref{sec:bh}, we start with a detailed description of
the disordered Bose-Hubbard model and the different disorder
realizations. After a short review of the phase diagram of the
disordered Bose-Hubbard model, we provide the definition of the overlap
function $q$ characterizing the disorder-induced correlations between
independent realizations of the system. In section~\ref{sec:meas}, we
describe in detail the preparation of the system and the subsequent
measurement scheme for the overlap function. In section~\ref{sec:overlap},
we determine the overlap for the disordered Bose-Hubbard model via
perturbation theory for physically-relevant limits and compare the
result with numerical simulations of the one-dimensional system.
Details of the calculations are presented in the appendices.

\section{Bose-Hubbard model with disorder}
\label{sec:bh}

Cold atomic gases subjected to an optical lattice are well described
by the Bose-Hubbard model \cite{jaksch:98,schulte:05}
\begin{equation}
   H= - J\sum_{ \langle i j \rangle} b_{i}^{\dag} b_{j} + \frac{U}{2} \sum_{i}
n_{i}(n_i-1)- \sum_{i} (\mu-\Delta_{i}) n_{i}, \label{dBHM}
\end{equation}
with $b^{\dag}_{i}$ ($b_{i}$) the bosonic creation (annihilation)
operator and $n_{i}= b_{i}^{\dag} b_{i}$ the particle number operator
at the lattice site $i$. The first term describes the kinetic energy
of the atoms with hopping energy $J$ between nearest-neighbour sites,
the second term accounts for the onsite interaction between the atoms,
and the third term describes the disorder potential with random-site
off-sets $\Delta_{i}$ of the chemical potential $\mu$.

The disorder potential $\{\Delta_{i}\}$ can be generated via
different methods \cite{lye:05,fort:05,clement:05,fallani:07,vignolo:03,gavish:05}
and is determined by a probability distribution $P(\delta)$ describing
the probability of having an onsite shift with strength $\delta$. The
mean square of the disorder distribution
\begin{equation}
\Delta^2 = \int d\delta \: \delta^2 P(\delta)
\end{equation}
gives rise to a characteristic energy scale $\Delta$ of the disorder
potential (note that the mean energy shift of the disorder potential is
absorbed in the definition of the chemical potential). In addition, the
disorder potential is characterized by a correlation length. Here, we
are interested in short-range disorder with the disorder in different
wells independent of each other.  The probability distribution $P(\delta)$ and its
correlation length depend on the source of the disorder potential, which
varies depending on the microscopic implementation in cold gases. The most
promising possibilities in order to produce disorder with short correlation
length involve the use of laser speckle patterns and mixtures of different cold atomic gases.
While first experiments with laser speckles had a disorder correlation
length larger than the lattice spacing \cite{lye:05,fort:05},
experimental efforts towards local addressability in an optical
lattice also offer the possibility for the creation of disorder with
a correlation length comparable to the lattice spacing, as well
as Gaussian probability distributions \cite{lee:07,gorshkov:07}.
Alternatively, a disorder potential can also be created in mixtures of
cold atomic gases in optical lattices \cite{guenter:06,ospelkaus:06}
by quenching the motion of one species. Then, the disorder correlation
length is of the order of the Wannier function, which can be smaller
than the lattice spacing due to the strong lattice that quenches the
motion of the disorder species, whilst the probability distribution
becomes bimodal with $\Delta_i = \pm \Delta$; here we are primarily
interested in such a short ranged disorder with $\Delta_i$ being
independent in the different wells.  Note that both
Gaussian disorder, as well as bimodal disorder generally give rise
to different physical phenomena in glass physics and thus both are
interesting in their own right.

\begin{figure}[h!]
\centerline{\includegraphics[width=8cm]{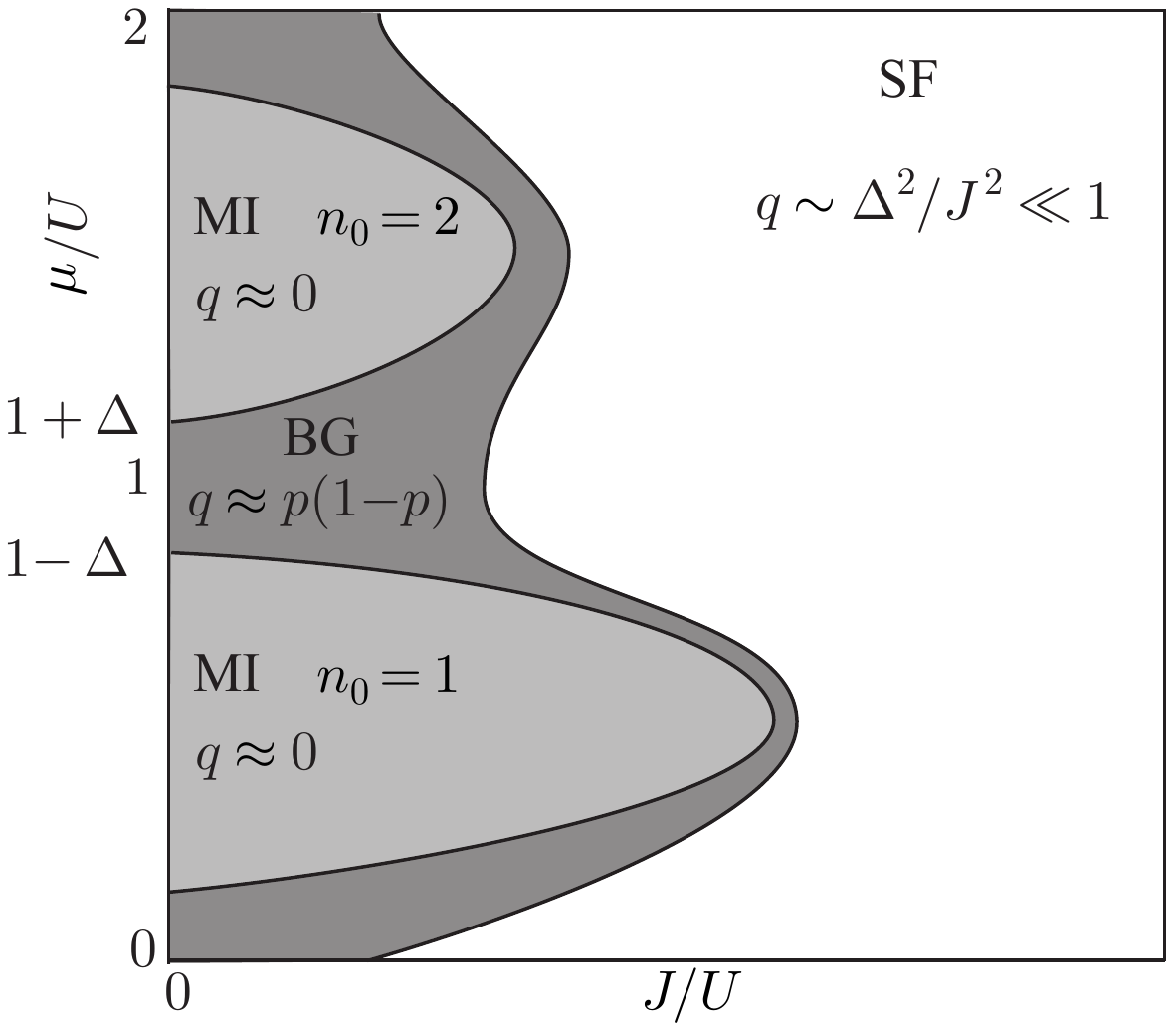}}
\caption{
Sketch of the phase diagram for the disordered Bose-Hubbard model:
the Mott-insulator (MI) appears for integer filling and the overlap
function $q$ vanishes for small hopping $J/U\ll 1$, while in the
Bose-glass phase (BG) the overlap function approaches a finite value
$q=p(1-p)$ with $p\in(0,1)$ characterizing the bimodal disorder
distribution (see section~\ref{sec:overlap}). Consequently
a measurement of the overlap allows for a clear distinction between
the Bose-glass and the Mott-insulator phase in this regime. In turn,
the superfluid phase (SF) is characterized by off-diagonal long-range
order resulting in coherence peaks in a time of flight experiment.
} \label{fig:phase_diagram}
\end{figure}

The zero-temperature phase diagram of the Hamiltonian
~(\ref{dBHM}) was first studied by Fisher {\em et
al.}~\cite{fisher:89} where three different phases were discussed:
the superfluid, the Mott-insulator, and the Bose-glass phases. In
the two-dimensional regime of interest here, the superfluid
phase appears for large hopping energies $J \gtrsim U, \Delta$
and is characterized by off-diagonal (quasi) long-range order
(finite superfluid condensate) and a linear excitation spectrum
giving rise to a finite compressibility. On the other hand,
for dominating interaction energies $U\gg \Delta , J$ the ground
state corresponds to an incompressible Mott-insulator phase with
an excitation gap and a commensurate filling factor, i.e., the
averaged particle number in a single well $\langle n_{i} \rangle
\in \mathrm{N}$ is an integer. While the quantum phase transition
from the superfluid to the Mott-insulator has been experimentally
identified \cite{greiner:02}, the disorder potential gives rise to an
additional Bose-glass phase characterized by a vanishing excitation
gap and finite compressibility; see figure~\ref{fig:phase_diagram}
for a sketch of the phase diagram.  The details of the phase diagram
have been studied via analytical methods in the regime of Anderson
localization \cite{gavish:05,kuhn:05,demartino:05,lugan:07b} and in other regimes
via numerical methods such as quantum Monte Carlo \cite{scalettar:91},
density matrix renormalization group (DMRG) \cite{rapsch:99},
dynamical mean-field theory \cite{byczuk:05}, and analytical methods
\cite{yukalov:07,krutitsky:08}. Furthermore, recent interests focused
also on the appearance of alternative phases for different disorder
types in the Bose-Hubbard model, e.g., off-diagonal disorder giving
rise to a Mott-glass phase \cite{sengupta:07} while random onsite interactions
can give rise to a Lifshits glass \cite{lugan:07a}.

The superfluid phase in the absence of disorder is experimentally
identified via a measurement of the coherence peaks characterizing
the condensate fraction, while the transition to a Mott-insulating
phase is characterized by the disappearance of these interference
patterns and as well as a change in the behaviour of excitations
\cite{greiner:02,stoeferle:04}.  However, the identification
of the Bose-glass phase in the presence of disorder requires an
additional observable allowing for the distinction between the
Mott-insulator and the Bose-glass, where the condensate fraction
vanishes. Such an additional property is known from spin-glass theory
as the Edwards-Anderson order parameter \cite{edwards:75,binder:86}
which appears as an order parameter in the mean-field theory on spin
glasses \cite{parisi:83}. In the present situation with a disordered
Bose-Hubbard model, its generalization leads to
\begin{equation}
  q_{\scriptscriptstyle EA} = [\left(\av{n_{i}} - n\right)^2]_{av} = [\av{n_{i}}^2]_{av}-
\left([\av{n_{i}}]_{av}\right)^2 \, ,\label{eq:q_EA}
\end{equation}
where $n = [\av{n}_{i}]_{av}$ denotes the mean particle density in
the sample. It is important to note the different averages involved
in the definition of the order parameter $q_{\scriptscriptstyle EA}$:
for a fixed disorder realization, the average $ \av{...}$ denotes
the thermodynamic average over the ground state of the system, while
$[...]_{av}$ describes the disorder average over different disorder
realizations. While the experimental determination of this quantity
in the Bose-Hubbard model requires an exact state tomography of the
system for each lattice site, a simpler and alternative route is
obtained by preparing two systems with the same disorder landscape
and measuring the correlations between these two decoupled systems:
the remaining correlation is disorder induced and disappears for weak
disorder. This so-called overlap function $q$ between the two systems
realized with the same disorder landscape takes the form
\begin{equation} \label{eq:q}
q = [\av{n_i^\alpha} \av{n_i^\beta}]_{av} -
[\av{n_i^\alpha}]_{av}[\av{n_i^\beta}]_{av}
\end{equation}
and where $\alpha$ and $\beta$ describe the two different systems
with the same disorder. This definition follows in close analogy of
the mean-field order parameter in the replica theory of spin glasses
\cite{parisi:83}.

\section{Measurement of the overlap function} \label{sect:q_measurement}
\label{sec:meas}

In this section we outline the key process of this work, i.e.,
the basic experimental procedure by which physical replicas can
be prepared and used to measure the overlap function $q$ defined
in~(\ref{eq:q}). This consists of three essential steps: (1)
initial preparation of disorder replicas, (2) introduction of probe
atoms and (3) the measurement process, involving recombination of the
replicas and spectroscopy to measure the overlap itself. At no stage
do we assume individual addressability of lattice sites, but rather
consider global operations and the measurement of global quantities.

\subsection{Initial preparation of disorder replicas}

We consider a situation in which atoms are confined to
a three-dimensional optical lattice, which is particularly deep
along the $z$-direction, so that an array of planes with 2D lattice
potentials are formed in the $x$--$y$ plane, with no tunneling
between the planes. Each of these 2D planes of the potential can be
divided into two subplanes, again along the $z$-direction, by adding
a superlattice of half the original period which can be controllably
switched on and off \cite{sebby-strabley:06,foelling:07}. The resulting
two subplanes then constitute our replicas, and our goal is that
these pieces should have the same disorder realisation.

In the case of disorder generated optically (using, e.g., speckle
patterns) the different layers exhibit the same disorder
landscape for a suitable orientation of the lasers.  However, for a
disorder induced by a second species, the replicas must be produced in
several steps.  We start with a two-species mixture in the undivided
plane. A fast ramping up of the optical lattice depth within the
plane for the disorder species results in a Poissonian distribution of
particles in the lattice. The next step is then a filtering procedure
removing singly-occupied sites and keeping only doubly-occupied and
unoccupied sites (which are distributed randomly). Such a filtering
procedure has been achieved in recent experiments where atoms are
combined to Feshbach molecules, and atoms remaining unpaired are
removed from the system using a combination of microwave and optical
fields that is energy-selective based on the molecular binding energy
\cite{winkler:06}. Finally, the superlattice is applied adiabatically
providing a double-layer structure, where exactly one atom of the
disorder species is transferred to each of the two subplanes. The
result is two planes with a random distribution of atoms of the
``disorder species,'' with an exact copy in each subplane, i.e.,
a replica of the random disorder pattern. This is illustrated in
figure~\ref{Fig:twolayers}.

Interactions between atoms of a probe species and the disorder species
can now be adiabatically increased, or the probe species otherwise
adiabatically introduced to the system (e.g., using spin-dependent
lattices or superlattices). This results in two replicas of the same
system, which we label $\alpha$ and $\beta$, with atoms in each
replica evolving independently according to their system Hamiltonian in
the same disorder potential.  We again reiterate that the superlattice
separating the replicas should be deep, so that there is no tunnelling
between the replicas. The only correlations between the two should
thus be determined by the identical disorder realisations.

\subsection{Measurement process}

\begin{figure}[h!]
\centerline{\includegraphics[width=8cm]{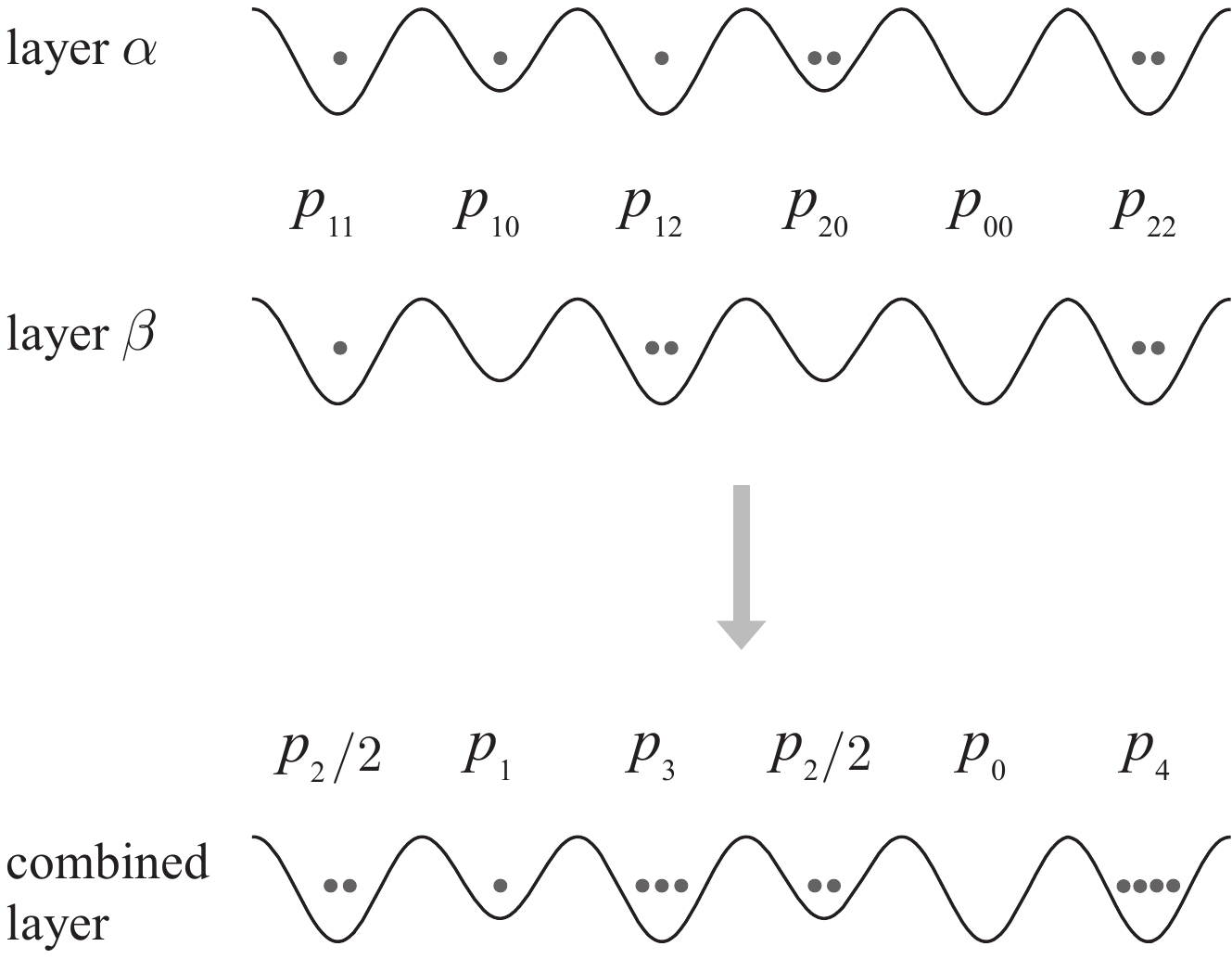}}
\caption{
A sample system illustrating the required joint and combined system
probabilities involved in the measurement scheme before and after
combining the two layers respectively. Top: The individual layers
$\alpha$ and $\beta$ and the corresponding joint probability $p_{nm}$
at each site. Bottom: Combined layer and the corresponding combined
system probability $p_n$ at each site. Note that this is only a guide,
because after combining the layers the particles are spread across
multiple motional states (not depicted).
} \label{fig:overview}
\end{figure}

To measure the overlap $q$ as given by~(\ref{eq:q}) we need
to compute correlation functions that compare the state of probe
atoms in the two replicas $\alpha$ and $\beta$. In each case,
the correlation functions can be computed if we measure the joint
probability distribution for occupation numbers in the two replicas
(this is discussed in detail below). In particular, we need to determine
the joint probability of having $n$ particles in layer $\alpha$ and $m$
particles in layer $\beta$ at a given lattice site, denoted by $p_{nm}$, and
the probability of having $n$ particles in layer $\alpha$, $\beta$ at a given
lattice site, denoted by $p_n^{\alpha,\beta}$ (both depicted in
figure~\ref{fig:overview}). We thus first freeze the state
in each replica by ramping the lattice suddenly to deeper values
in all three directions, so that tunnelling no longer occurs. We
can then measure the different possible configurations of particles
occupying individual lattice sites in the layers $\alpha$ and $\beta$
(see figures~\ref{fig:overview}--\ref{fig:species_transfer_n_2}).

This is done in two steps: First, the layers are combined so that at
each site each initial eigenstate involving different occupation
numbers in the two replicas becomes a superposition of degenerate
states involving the total number of atoms spread over motional states
in a single well (see figure~\ref{fig:overview_combining_layers},
section~\ref{sect:comb_layers} below). The final state is, of course,
strongly dependent on the combination process, as well as the initial state before the combination
of the layers. However, we need only determine the fraction
of lattice sites with a total of $n$ particles, denoted $p_n$, which is unchanged provided that the
lattice is deep within each plane, to prevent tunnelling of atoms. The
relative frequency with which each final configuration $p_n$ occurs
can then be measured based on the different spectroscopic energy
shifts arising from different numbers of atoms in the final well,
and different distributions of atoms over the motional states (see
figure~\ref{fig:species_transfer_n_2}, section~\ref{sect:species_transfer}
below). This involves weak coupling of atoms to an auxiliary internal
state, similar to that performed in \cite{campbell:06a}.
We show that the set of possible final states given an initial
configuration can be distinguished spectroscopically from the set
of states corresponding to initial configurations giving different
contributions to $p_{n}$, and thus we can reconstruct the original
joint probability distribution $p_{nm}$ prior to combining the
layers. The reduced probabilities $p_n^{\alpha,\beta}$ can be
determined from similar spectroscopy without first combining the
layers, allowing the full overlap function $q$ to be constructed. Below
we give further details of each step in this procedure.

\subsubsection{Quantities that need to be measured}

We need to measure two quantities to determine the overlap $q$,
namely $[\av{\hat{q}_1}]_{av}$, where
\begin{equation}
\hat{q}_1 = \frac{1}{L}\sum_i n_i^\alpha n_i^{\beta}
\end{equation}
with $\alpha \neq \beta$ and $L$ is the total number of lattice
sites, corresponding to the density-density correlations
between the two layers, and $[\av{\hat{q}^\alpha_2}]_{av}$,
$[\av{\hat{q}^\beta_2}]_{av}$, where
\begin{equation}
\hat{q}_2^{\alpha,\beta} = \frac{1}{L} \sum_i n_i^{\alpha,\beta},
\end{equation}
corresponding to the average density of each layer. Then we can express
\begin{equation}
q=[\av{\hat{q}_1}]_{av}-[\av{\hat{q}^\alpha_2}]_{av}
[\av{\hat{q}^\beta_2}]_{av}.
\end{equation}
The disorder average should be performed by repeated measurements of
the quantum average for different realisations of the disorder \cite{comment:ciracparallel}. In
general, one can expect that a measurement on $n_{i}$ and $n_{j}$
for large separation between sites $i$ and $j$ is independent of each
other, and consequently the summation over lattice sites automatically
performs a statistical quantum average \cite{comment:stats}.
In the following, we are interested  in small filling factors such
that lattice sites with three or more particles are rare and can be
neglected in all three phases of the system.

First we focus on measurement of
$[\av{\hat{q}_2^\alpha}]_{av}[\av{\hat{q}_2^\beta}]_{av}$,
i.e., the second term of the overlap in~(\ref{eq:q}). If
the layers are prepared as discussed above, the symmetry
$(1/L)\sum_i\av{n_i^\alpha} = (1/L)\sum_i\av{n_i^\beta}$
is preserved and thus we can measure the average over two replicas,
$[\av{\hat{q}_2}]^2$ where \cite{comment:occupation}
\begin{equation}
\hat{q}_2= \frac{1}{2}(\hat{q}_2^\alpha
+ \hat{q}_2^\beta) =  \frac{1}{2L} \sum_i (n_i^{\alpha}+n_i^\beta) .
\end{equation}
For the average density we obtain:
\begin{equation}
\av{\hat{q}_2} = \frac{1}{2L}\sum_i\av{n_i^\alpha} + \av{n_i^\beta} =
\frac{1}{2}\sum_{n} n (p_n^\alpha+p_n^\beta) ,
\end{equation}
where $p_n^{\alpha,\beta} = N_n^{\alpha,\beta}/L$ with
$N_n^{\alpha,\beta}$ the number of sites with $n$ particles in layers
$\alpha,\beta$, and thus
\begin{equation}
\av{\hat{q}_2}= \frac{1}{2}(p_1^{\alpha}+p_1^{\beta} + 2 p_2^{\alpha}
+ 2 p_2^{\beta}) .
\label{eq:q_second_part}
\end{equation}

Next we consider the measurement of $[\av{\hat{q}_1}]$
i.e., the first part of the overlap in~(\ref{eq:q}). In
this case the density-density correlations can be expressed as
\begin{equation}
\frac{1}{L}\sum_i\av{n_i^\alpha n_i^\beta}= \sum_{n} n m p_{nm} ,
\end{equation}
where $p_{nm} = N_{nm}/L$ with $N_{nm}$ the total number of sites with
$n$ particles in layer $\alpha$ and $m$ particles in layer $\beta$,
and thus, keeping terms up to two particles in total over the replicas
at a given site
\begin{equation}
\label{density_density_correlations}
\av{\hat{q}_1} = \frac{1}{L}\sum_i\av{n_i^\alpha n_i^\beta} = p_{11} +
2(p_{12}+p_{21}) + 4p_{22}.
\end{equation}
Now consider $p_n = N_n/L$, with $N_n$ the total number of
particles from both layers, which can be measured by combining
both layers (described in detail in section~\ref{sect:comb_layers})
and subsequently applying a similar scheme to the one
presented in \cite{campbell:06a} (described in detail in
section~\ref{sect:species_transfer}). Using the following equations that
relate both $p_n$ and $p_n^{\alpha,\beta}$ to $p_{nm}$
\begin{displaymath}
\begin{array}{llllll}
p_1 &=& p_{01} + p_{10} \qquad &p_1^\alpha &=& p_{10} + p_{12} + p_{11},\\
p_2 &=& p_{20} + p_{02} + p_{11} \qquad &p_1^\beta &=&  p_{01} + p_{21} +
p_{11},\\
p_3 &=& p_{12} + p_{21} \qquad & p_2^\alpha &=& p_{20} + p_{21} + p_{22},\\
p_4 &=& p_{22} \qquad &p_2^\beta  &=&  p_{02} + p_{12} + p_{22},
\end{array}
\end{displaymath}
we can show that $p_{11} = \frac{1}{2} (p_1^\alpha + p_1^\beta
-p_1 -p_3)$ which in conjunction with $p_3=p_{12}+p_{21}$
and $p_4=p_{22}$ gives all necessary terms. Then,
~(\ref{density_density_correlations}) simplifies to
\begin{eqnarray}
\av{\hat{q}_1} =  \frac{1}{2} (p_1^\alpha + p_1^\beta -p_1) +
\frac{3}{2} p_{3} + 4p_{4}.
\end{eqnarray}
Finally, note that $\hat{q}_2$ as given in
~(\ref{eq:q_second_part}), can also be expressed as $\av{\hat{q}_2}
= \frac{1}{2}\sum_n n p_n$.

In the following two sections we show how to measure these
conditional probabilities by combining the two replicas together,
and performing spectroscopy where a single atom is coupled to a
different internal state. In section~\ref{sect:comb_layers} we describe
the process of combining the two layers into a single one whilst in
section~\ref{sect:species_transfer} we describe a similar scheme to
the one presented in \cite{campbell:06a} that can be used to
distinguish different occupation numbers.

\subsubsection{Combining Layers} \label{sect:comb_layers}

We now describe in more detail the process of combining the two
layers into a single one and determine the resulting joint state
for the different possible initial configurations. The combination
of the layers is achieved by lowering the potential barrier that
separates both layers (switching off the superlattice). Specifically,
we now determine the final (after completely lowering the barrier)
state for each of the different initial configurations of particles
(prior to lowering the barrier).

After the superlattice is removed and layers are combined, the system
is described by the Hamiltonian
\begin{eqnarray} \label{H_int_SA}
H_{\rs comb} &=& \sum_{k\in \{0,1\} }\left[ \epsilon_{k}n_k +\frac{U_{kk}}{2}n_k(n_k-1)\right] + 2U_{01}n_0n_1, \nonumber\\
\end{eqnarray}
where $U_{kl} = 2\omega_\perp a \int dz |w_k(z)|^2|w_l(z)|^2$,
with $a$ the scattering length, $\omega_\perp$ the transverse
trapping frequency, $w_k(z)$ the Wannier functions for band $k$,
and $\epsilon_k$ the band energy ($k,l\in \{ 0,1 \}$). The expression for $U_{kl}$
is a special case of the general 3D expression
\begin{eqnarray}
U = \frac{4 \pi \hbar^2 a}{m} \int dx dy dz |w_0(x)|^4 |w_0(y)|^4 |w_k(z)|^2|w_l(z)|^2, \nonumber
\end{eqnarray}
where $w_0(x)$ and $w_0(y)$ are the Wannier functions for the lowest motional state in the 2D replica plane $(x,y)$,
with the additional assumption of an isotropic and deep lattice potential in the replica plane characterized by the frequency $\omega_\perp$.

\begin{figure}
\centerline{\includegraphics[width=8cm]{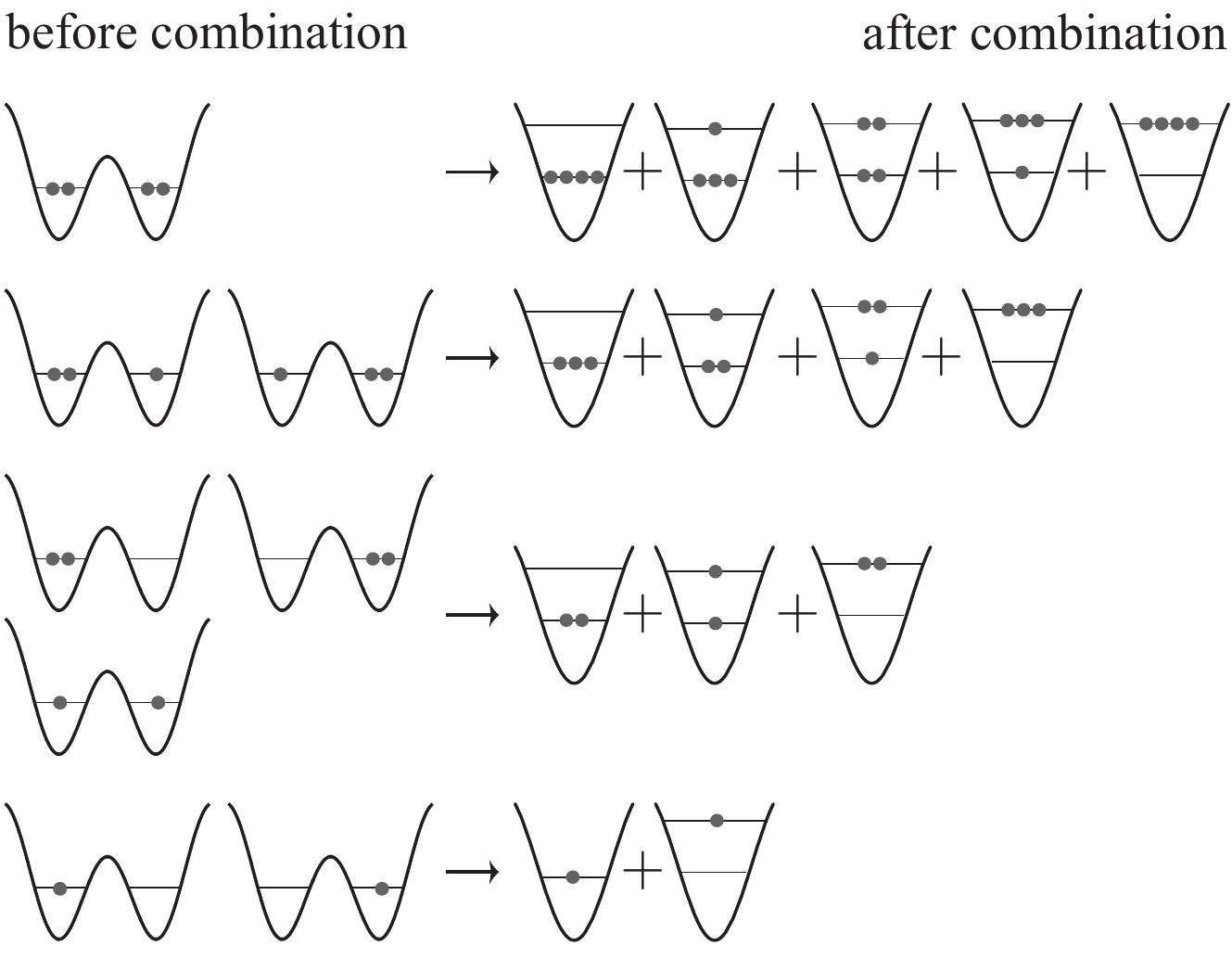}}
\caption{
Analysis of the process of combining the two initial replicas,
showing possible states with at most double occupation (dots) in a
given well before combination of the layers (left hand side), and the
resulting possible states after combination of the layers (right hand
side). Note that here `$+$' indicates a general superposition with
unspecified coefficients, as these are dependent on the precise time
dependence of system dynamics during combination. Knowledge of these
coefficients is not necessary in the measurement scheme presented here.
} \label{fig:overview_combining_layers}
\end{figure}

States which are initially nondegenerate will evolve
adiabatically to the corresponding eigenstate of the Hamiltonian
in~(\ref{H_int_SA}), provided the barrier is lowered slowly
with respect to the energy separation between the (nondegenerate)
eigenstates. However, we note that some of the initial states are
doubly degenerate due to the equal single particle energies in
the two layers. Thus when the barrier is lowered these degenerate
states become coupled to each other (because the single particle
states change time dependently) and the final state will evolve into
a process-dependent superposition of the different corresponding
eigenstates of the Hamiltonian in~(\ref{H_int_SA}). However, this
poses no problem for our measurement scheme since the states that
are initially degenerate contribute equally to the probabilities we
are trying to measure. Moreover, for a given initial configuration
it is not necessary to know the full final state, in fact it is
sufficient to know only the possible corresponding final eigenstates
of the Hamiltonian in~(\ref{H_int_SA}) (with the same number of
particles as the initial configuration) that contribute to the final
state. Figure~\ref{fig:overview_combining_layers} depicts all the
different possible states before and after combining the layers for
up to a maximum of four particles in total.

In the next section we show how the different configurations for a
given particle number can be measured and thus the total occupation
numbers determined.

\subsubsection{Number Distribution Characterization}
\label{sect:species_transfer}

We now investigate a scheme similar to the one presented in
\cite{campbell:06a} that uses density-dependent energy shifts to
spectroscopically distinguish different particle numbers at each site
of the combined layer (see section~\ref{sect:comb_layers} above) resulting
in a measurement of $p_n$. The measurement of $p_n^{\alpha,\beta}$
follows from the same spectroscopic analysis, but performed on each
individual layer $\alpha$, $\beta$ before combining the layers.
Such a scheme has already been implemented experimentally and was
utilized to observed the Mott-insulating shell structure that arises
due to a harmonic trapping potential commonly present in
cold atomic gases experiments \cite{campbell:06a}.

%We now investigate a scheme similar to the one presented in
%\cite{campbell:06a} that uses density-dependent energy shifts to
%spectroscopically distinguish different particle numbers at each site
%of the combined layer (see section~\ref{sect:comb_layers} above).

We begin with all atoms in an internal state $a$, and investigate the
weak coupling of particles to a second internal state $b$ (with at
most one particle transferred). Due to the difference in interaction
energy between particles of species $a$ and $b$, and particles solely
of species $a$ there is an energy shift between the ``initial''
state (all particles of species $a$) and ``final'' state (one
particle of species $b$ and all others particles of species $a$). Thus
different initial particle numbers can be distinguished by the energy
shift (corresponding to the Raman detuning at which the transfer is
resonant), provided the shift is different for the different particle
numbers. Since particles in the combined layer can be spread across
multiple motional states (see section~\ref{sect:comb_layers} above),
different energy shifts occur for different configurations of the
same number of particles. Whilst we find that the energy shifts
of some of the configurations (of the same number of particles)
are identical or very similar, there are other configurations
having substantially different energy shifts. Thus we have to check
that the energy shifts for all configurations of a given number of
particles can be distinguished from energy shifts of configurations
corresponding to a different total number of particles. To this end
we now present a detailed calculation of the energy shifts arising
from all the different initial configurations of particles and show
in which cases the total number of particles can be distinguished.

Let $a_k^\dagger$ and $b_k^\dagger$ denote the creation operators for particles
of species $a$ and $b$, respectively, in band $k$ where $k \in \{
0,1\}$. Then the Raman coupling between the two internal states $a$ and
$b$ is described by an effective Hamiltonian (within a rotating-wave
approximation)
\begin{equation}
H_{\rs RC} = \frac{\Omega(t)}{2} (a^\dagger_0 b_0 + a^\dagger_1 b_1 +
\textrm{H.c.}) - \Delta(t) (b_0^\dagger b_0+ b_1^\dagger b_1),
\label{raman_coupling}
\end{equation}
where $\Delta(t)$ is the Raman detuning, $\Omega(t)$ the effective
two-photon Rabi frequency, and we have assumed that $\Omega(t)\ll
\Delta'$ where $\Delta'$ are the detunings of the lasers from the
atomic excited state. In what follows we assume that $\int \Omega(t)
dt \ll \pi$, i.e., weak coupling to internal state $b$ and that the two
lasers creating the Raman transition are running waves with equal wave
vectors. Note that processes in which the internal state and the band
would change do not occur provided the Raman detuning and effective
two-photon Rabi frequency are smaller than the band separation.

The onsite interaction energy for atoms of species $a$ and $b$ in
the lowest two bands is described by the following Hamiltonian
\begin{eqnarray}
H_{\rs int} & = & \sum_{i\in \{a,b\}} \bigg[\frac{U_{00}^{ii}}{2} n_0^i(n_0^i-1)
+ \frac{U_{11}^{ii}}{2} n_1^i(n_1^i-1) \nonumber \\
&& + 2U_{01}^{ii} n_0^in_1^i \bigg] + U_{00}^{ab} n_0^an_0^b + U_{11}^{ab} n_1^a
n_1^b \nonumber + U_{01}^{ab} \left[ n_0^a n_1^b \right .\\
&& \left. +n_1^a n_0^b + (a_0^\dagger a_1 b_1^\dagger b_0 + a_1^\dagger a_0
b_0^\dagger b_1 + \textrm{H.c.}) \right], \nonumber \\
\label{hamiltonian_two_bands_two_species}
\end{eqnarray}
where
\begin{equation}
U_{kl}^{ij} = 2 \omega_\perp a_{ij} \int dz
|w_k(z)|^2|w_l(z)|^2,
\end{equation}
with $a_{ij}$ the scattering length between two particles in internal
states $i$ and $j$ ($i,j \in \{ a,b \}$), $\omega_\perp$ the transverse
trapping frequency, and $w_k(z)$ the Wannier functions corresponding
to band $k$ ($k,l\in \{0,1 \}$). Note we have also introduced the
particle number operators $n_k^a = a_k^\dagger a_k$ and $n_k^b =
b_k^\dagger b_k$. We further make the assumption that the lattice
is very deep so that we can approximate all the Wannier functions by
simple harmonic oscillator eigenfunctions. This assumption of a deep
lattice gives the following simplified expressions $2U_{01}^{ij} =
U_{00}^{ij}$ and $U_{11}^{ij} = (3/4) U_{00}^{ij}$.

Using~(\ref{hamiltonian_two_bands_two_species}) we now explicitly
compute the energy shift $\Delta E = E_f - E_i$ associated with the
processes $\ket{i} \rightarrow \ket{f}$, where $\ket{i}$ is the initial
state corresponding to all particles of species $a$ and having energy
$E_i$, and $\ket{f}$ is the final state corresponding to one particle
transferred to internal state $b$ and having energy $E_f$.  Clearly,
$\Delta E$ depends on how the particles are initially distributed in
the lower and higher bands after combining the two layers (see the
right hand side of figure~\ref{fig:overview_combining_layers}). There
are two different cases to be considered; either all particles
are initially in the same band (this is the situation in
\cite{campbell:06a}) or particles are initially in both bands. In
the latter case it is then possible that particles in either band
are transferred to internal state $b$ and since these different
possible resulting states are coupled by the interaction Hamiltonian
in~(\ref{hamiltonian_two_bands_two_species}) the final resulting
state is a superposition of these, corresponding to an eigenstate of
the Hamiltonian in~(\ref{hamiltonian_two_bands_two_species}).

\begin{figure}[h!]
\centerline{\includegraphics[width=8cm]{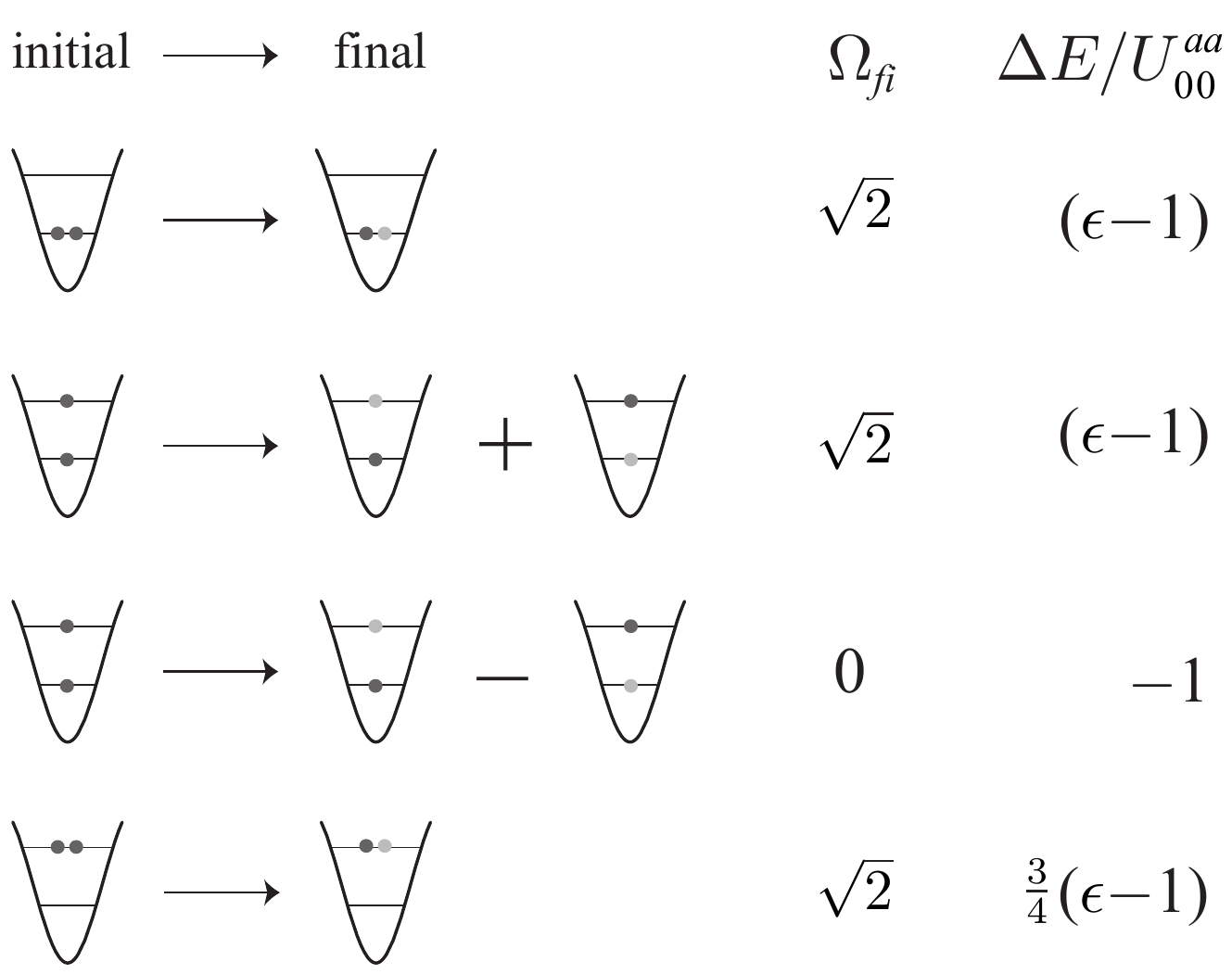}}
\caption{
Illustration of the number distribution characterization within the
two layers for states with two particles in the combined layer. On
the left are all possible initial configurations of two particles
of species $a$ in the combined lattice (dark-coloured circles) and the
corresponding final states with one particle each of species $a$
and $b$ (light-coloured circles). On the right we list the respective
coupling matrix elements, $\Omega_{fi}$, and energy shifts, $\Delta
E /U_{00}^{aa}$, for each of the configurations on the left.
}
\label{fig:species_transfer_n_2}
\end{figure}

For example, let us consider a total initial number of two particles as
depicted in figure~\ref{fig:species_transfer_n_2}. The different initial
and final states are most conveniently expressed in the occupation
number basis $\ket{N_0^a,N_1^a;N_0^b,N_1^b}$ with $N_0^{a,b}$
($N_1^{a,b}$) the number of particles of species $a$, $b$ in the lower
(higher) band at a given lattice site. First consider the case where
both particles are initially in the lower or higher bands corresponding
to the initial states $\ket{2,0;0,0}$ or $\ket{0,2;0,0}$ (see top and
bottom of figure~\ref{fig:species_transfer_n_2}). The corresponding
final states are given by $\ket{1,0;1,0}$ and $\ket{0,1;0,1}$,
and the corresponding energy shifts are $\Delta E/U_{00}^{aa} =
\epsilon-1$ and $\Delta E/U_{00}^{aa} = \frac{3}{4}(\epsilon-1)$,
where we have defined $\epsilon \equiv U_{00}^{ab}/U_{00}^{aa}$. Next
we consider the case where there is a particle in each band
initially corresponding to the initial state $\ket{1,1;0,0}$
(see middle of figure~\ref{fig:species_transfer_n_2}). Then
the corresponding final states are the eigenstates of the
Hamiltonian in~(\ref{hamiltonian_two_bands_two_species})
in the subspace corresponding to one particle of species $a$
and one particle of species $b$ given by $\ket{1,1}_\pm \equiv
(1/\sqrt{2})\left(\ket{0,1;1,0} \pm \ket{1,0;0,1}\right)$, with
eigenenergies $E_+/U_{00}^{aa} = \epsilon$ and $E_-/U_{00}^{aa}=0
$. The associated energy shifts for these two possible final states
are $\Delta E /U_{00}^{aa} = \epsilon-1$ and $\Delta E /U_{00}^{aa}
= -1$ respectively. Note that the Raman transition only couples the
state $\ket{1,1}_+$ to the initial state  $\ket{1,1;0,0}$ with a
matrix coupling element $\Omega_{fi} = \sqrt{2}$, where $\Omega_{fi}
\equiv \bra{f} H_{\rs RC} \ket{i}$, while for the state $\ket{1,1}_-
$ we find $\Omega_{fi}=0$. Figure~\ref{fig:species_transfer_n_2}
summarizes the different possible configurations for a total of two
particles together with the corresponding energy shift $\Delta E$
and matrix coupling element $\Omega_{fi}$. We note that the energy
shifts, $\Delta E$, for the different final states to which the
initial states couple (i.e., $\Omega_{fi} \neq 0$) are very similar,
this is of advantage since our scheme needs only to distinguish energy
shifts for different {\em total} particle numbers.

Similarly, we can determine the energy shifts and coupling matrix
elements for other numbers of particles. If all particles are
initially in the lower or higher state corresponding to the
initial states $\ket{N_0^a,0;0,0}$ or $\ket{0,N_1^a;0,0}$, the
corresponding final states are given by $\ket{N_0^a-1,0;1,0}$
and $\ket{0,N_1^a-1;0,1}$. The corresponding energy shifts are
then given by $\Delta E /U_{00}^{aa} = (N_0^a-1)(\epsilon-1)$
and $\Delta E /U_{00}^{aa} = \frac{3}{4}(N_1^a-1)(\epsilon-1)$,
and the matrix coupling element in each case is $\Omega_{fi} =
\sqrt{N_0^a}$ and $\Omega_{fi} = \sqrt{N_1^a}$ respectively. For
other initial configurations with $m$ particles in the lower
band and $n$ particles in the higher band the corresponding
final states are found by diagonalizing the Hamiltonian in
~(\ref{hamiltonian_two_bands_two_species}) in the appropriate
particle subspace. This always results in a pair of eigenstates
denoted $\ket{m,n}_\pm$ with eigenenergy $E_\pm^{(m,n)}$ and using
these expressions the energy shift $\Delta E$ and coupling matrix
elements $\Omega_{fi}$ can be directly determined. For a total initial
number of three and four particles the expressions for $\ket{m,n}_\pm$,
$E_\pm^{(m,n)}$, $\Delta E$ and $\Omega_{fi}$ are given in appendix
~\ref{appendix_1a} and~\ref{appendix_1b}, respectively, for all the
different initial configurations. In table~\ref{table:species_transfer}
we list the results for up to a maximum number of four particles;
the first column lists the different initial configurations and
corresponding final states, the second column lists the corresponding
matrix coupling elements $\Omega_{fi}$ and the third column lists
the associated energy shifts $\Delta E$.

It is possible to distinguish different occupation numbers for specific
ranges of $\epsilon$ if the energy shifts for different total numbers
of particles are distinct. The desired range of $\epsilon$ is then
between points where any energy shifts for different total number of
particles would coincide (e.g., $\epsilon =1$) \cite{comment:table}.
For a specific example we consider the value $\epsilon  = 0.98$
(corresponding to scattering length values in \cite{campbell:06a})
and give numerical values for the matrix coupling elements
$\Omega_{fi}$ and the energy shifts $\Delta E$ in columns four and
five, respectively, in table~\ref{table:species_transfer}.  Then,
to a high accuracy, the determination of $p_{n}$ is obtained by
applying the coupling between the different internal states $a$
and $b$ for a short time $t_{n}=\Delta t/\sqrt{n}$ at all resonant
frequencies in table~\ref{table:species_transfer} with total number
of particles $n$ (ignoring the resonances with small and vanishing
couplings $\Omega_{fi}$). The total number of transferred particles
is then proportional to $p_n$ and efficiently avoids problems arising
from non-adiabatic combination of the layers.

\begin{table}
\caption{
Resonant energies and coupling strengths for the transfer of single
atoms in the combined layer to a different internal state. Column
one lists the different initial configurations and the corresponding
final states up to a maximum number of four particles. In the second
and third column we list the matrix coupling elements $\Omega_{fi}$
and energy shifts $\Delta E$, respectively, corresponding to the
different initial configurations in the first column. The states
$\ket{m,n}_\pm$ are listed in the text and appendices together with
the corresponding eigenenergies $E_\pm^{(m,n)}$ and coupling matrix
elements $\eta_\pm^{(m,n)}$. The last two columns give numerical values
of $\Omega_{fi}$ and $\Delta E$ for the specific value $\epsilon=0.98$.
\label{table:species_transfer}}
\begin{tabular}{|l||r|r||r|r|}
\hline
\hline
$\ket{i} \rightarrow \ket{f}$ & $\Omega_{fi}$ & $\Delta E/U_{00}^{aa}$ & $\Omega_{fi}$ & $\Delta E/U_{00}^{aa}$ \\
\hline \hline
$\ket{1,0;0,0} \rightarrow \ket{0,0;1,0}$ & 1 & 0 & 1 & 0 \\
\hline
$\ket{0,1;0,0} \rightarrow \ket{0,0;0,1}$ & 1 & 0 & 1 & 0\\
\hline
$\ket{2,0;0,0} \rightarrow \ket{1,0;1,0}$ & $\sqrt{2}$ &
$\frac{3}{4}(\epsilon-1)$ & 1.4 & -.018 \\
\hline
$\ket{2,0;0,0} \rightarrow \ket{1,0;1,0}$ & $\sqrt{2}$ & $(\epsilon-1)$ & 1.4 & -.024 \\
\hline
$\ket{1,1;0,0} \rightarrow \ket{1,1}_+ $& $\sqrt{2}$ &  $(\epsilon-1)$ & 1.4 & -.024 \\
\hline
$\ket{1,1;0,0} \rightarrow \ket{1,1}_-$ & 0 & -1 & 0 & -1\\
\hline
$\ket{3,0;0,0} \rightarrow \ket{2,0;1,0}$  & $\sqrt{3}$ & $2(\epsilon-1)$ & 1.7 & -.049 \\
\hline
$\ket{2,1;0,0} \rightarrow \ket{2,1}_+$ & $\sqrt{3}$ & $2(\epsilon-1)$ & 1.7 & -.049 \\
\hline
$\ket{2,1;0,0} \rightarrow \ket{2,1}_-$  & 0 & $\frac{\epsilon}{2} -2$ & 0 & -1.5 \\
\hline
$\ket{1,2;0,0} \rightarrow \ket{1,2}_+$ & $\eta_+^{(1,2)}$ & $E_+^{(1,2)} - 3$ & 1.7 & -.29 \\
\hline
$\ket{1,2;0,0} \rightarrow \ket{1,2}_-$ & $\eta_-^{(1,2)}$ & $E_-^{(1,2)} - 3$ & -.0034 & -1.8 \\
\hline
$\ket{0,3;0,0} \rightarrow \ket{0,2;0,1}$ & $\sqrt{3}$ & $\frac{3}{2} (\epsilon-1)$ & 1.7 & -.037\\
\hline
$\ket{4,0;0,0} \rightarrow \ket{3,0;1,0}$ & $\sqrt{4}$ & $3 (\epsilon-1)$ & 2 & -.073\\
\hline
$\ket{3,1;0,0} \rightarrow \ket{3,1}_+$ & $\sqrt{4}$ & $3(\epsilon -1)$ & 2 & -.073 \\
\hline
$\ket{3,1;0,0} \rightarrow \ket{3,1}_-$ & 0 & $\epsilon-3$ & 0 & -2.0\\
\hline
$\ket{2,2;0,0} \rightarrow \ket{2,2}_+$ & $\eta_+^{(2,2)}$ & $E_+^{(2,2)} -\frac{23}{4}$ & 2.0 & -0.070 \\
\hline
$\ket{2,2;0,0} \rightarrow \ket{2,2}_-$ & $\eta_-^{(2,2)}$ & $E_-^{(2,2)} - \frac{23}{4}$ & -.0031 & -2.0 \\
\hline
$\ket{1,3;0,0} \rightarrow \ket{1,3}_+$ & $\eta_+^{(1,3)}$ & $E_+^{(1,3)} -\frac{15}{4} $ & 2.0 & 1.4\\
\hline
$\ket{1,3;0,0} \rightarrow \ket{1,3}_-$ & $\eta_-^{(1,3)}$ & $E_-^{(1,3)} -\frac{15}{4} $ & -.0054 & -.52\\
\hline
$\ket{0,4;0,0} \rightarrow \ket{0,3;0,1}$ & $\sqrt{4}$ & $\frac{9}{4}(\epsilon-1)$ & 2 & -.055\\
\hline
\hline
\end{tabular}
\end{table}

\section{Overlap Function in the Disordered Bose-Hubbard model}
\label{sec:overlap}

In this section we determine the overlap function $q$ [see~(\ref{eq:q})]
in the disorder Bose-Hubbard model, both analytically and
numerically. We focus on a bimodal disorder potential $\Delta_{i}=
\pm \Delta$ with the probability distribution
$P(\Delta)=1-P(-\Delta)= p$, which is the appropriate description
for the chosen disorder implementation (see discussion in earlier
sections). In the following, we present the modifications due to the
presence of a weak ($\Delta < U/2$) bimodal disorder potential, and
the prediction for the overlap function within the different
phases (see figure~\ref{fig:phase_diagram}). It follows from this
constraint that we do not enter the compressible disordered phase,
and hence the system will always have a unique ground state. Thus,
the ground states in two exact copies $\alpha$, $\beta$ of the system
(i.e. same disorder distribution) are the same, and consequently
$\langle n_i^{\alpha}\rangle=\langle n_i^{\beta}\rangle$. We
therefore expect a Sherrington-Kirkpatrick-type behaviour for the
overlap~\cite{sherrington:75}, i.e. the overlap between any
two copies will always be the same, and thus the overlap function $q$
can in fact be calculated using~(\ref{eq:q_EA}).

While the above is a self-evident statement for a replica-symmetric
system such as the one we examine here, it may no longer be true for
systems thought to exhibit replica-symmetry breaking (e.g. a
classical spin glass type model \cite{parisi:83}, or quantum systems
with extended interactions~\cite{giamarchi:01}). For the latter type
of system, measurement of the overlap function using physical replicas
(see section~\ref{sect:q_measurement}) is thought to yield different
values for the overlap depending on $\alpha$ and $\beta$, the
defining signature of replica symmetry-breaking.

In section~\ref{sect:analytical_particle_overlap} we analytically
calculate the overlap function, for the various phases of the disorder
Bose-Hubbard model~(\ref{dBHM}), whilst in section
~\ref{sect:numerical_particle_overlap} we present results for the
overlap function from numerical simulation of the one-dimensional system.

\subsection{Analytical determination of the overlap function}
\label{sect:analytical_particle_overlap}

Starting with the limit of vanishing hopping $J=0$, (\ref{dBHM})
reduces to an onsite Hamiltonian and the ground state is given by
$|\Omega\rangle =\Pi_{i} |\Omega\rangle_{i}$ with $|\Omega\rangle_{i}$
the lowest energy state within each site $i$. This lowest energy states
takes the form $|\Omega\rangle_{i}= |n_{i}\rangle$ with $|n_{i}\rangle$
being a state with fixed particle number $n_{i}$ at site $i$, with
$n_{i}$ subject to the constraint
\begin{equation}
 U \left( n_{i} -1 \right) < \mu - \Delta_{i} < U n_{i}.
\label{constraint}
\end{equation}
For a weak bimodal disorder potential ($\Delta < U/2$), we have to
distinguish two different cases.

First, for a chemical potential within the range $U (n_{0}-1)+\Delta <
\mu < U n_{0}-\Delta$ the above constraint in~(\ref{constraint})
is fulfilled independently of the site parameter $i$ for the
integer value $n_{0}$, i.e., at each site the ground state
is characterized by the integer particle density $[\langle n_i
\rangle]_{av} =n_{0}$ and we obtain a Mott-insulator (MI) phase (see
figure~\ref{fig:phase_diagram}). Furthermore, the overlap parameter
vanishes identically in this limit, i.e., $q=0$  for a Mott-insulator
at $J=0$.

Second, for chemical potentials in the range $U n_{0}-\Delta < \mu <
U n_{0}+\Delta$, the particle number at sites with different disorder
potential $\Delta_{i}$ differs by a single particle, i.e., at each
site with disorder potential $\Delta_{i}=\Delta$ the particle number
is determined by $n_{0}$, while at sites with disorder potential
$\Delta_{i}= -\Delta$ the lowest energy state is characterized by
$n_{0}+1$ particles. Therefore, the averaged particle density and
the overlap parameter take the form
\begin{equation}
 [\langle n_i \rangle]_{av} = n_{0} + (1-p)  \hspace{1cm}  q = p (1- p) .
\end{equation}
This situation corresponds to the disorder-induced Bose-glass (BG)
phase (see figure~\ref{fig:phase_diagram}) and the ground state is
characterized by nonhomogeneous filling. Similarly to the case of the
MI phase, the particle number is fixed over a finite range of the
chemical potential and thus the BG is incompressible. In contrast
to the MI phase, the BG phase is characterized by a non-vanishing
overlap parameter dependent on $p$.

Next, we focus on small hopping, $J\ll U,\Delta,\mu$, and determine
the overlap function using perturbation theory. We seek a unitary
transformation of the form $U=e^{-iS}$ which transforms the
total Hamiltonian in~(\ref{dBHM}) to an effective Hamiltonian
having the same eigenspectrum but acting only in the space of the
unperturbed eigenstates. Using this unitary transformation we can
directly determine the corrections due to hopping in perturbation
theory by computing $\bra{\Omega} \tilde{n}_i \ket{\Omega}$ where
$\tilde{n}_i \equiv U^\dagger n_i U$. Details are given in appendix
~\ref{app:pert_theory}.

We start with the Mott-insulator and consider the leading order
(non-vanishing) correction to the overlap function which occurs
at fourth order due to a site (disorder) independent density,
$\bra{\Omega} n_i\ket{\Omega}=n_0$ (using $\ket{\Omega}_i= \ket{n_0}$
in the MI). To lowest order in perturbation theory the overlap is
then given by
\begin{equation}
q  = J^4 \left([\av{\tilde{A}_2}^2]_{av} - [\av{\tilde{A}_2}]_{av}^2 \right),
\end{equation}
where
\begin{equation}
\tilde{A}_2 = -\frac{1}{2} [S_1,[S_1,n_i]] ,
\end{equation}
with
\begin{equation}
S_1= -i\sum_{\av{jk}} c_{jk} \ket{\Omega}\bra{\phi_{jk}} + \textrm{H.c.}
\end{equation}
and $\ket{\phi_{jk}}= \ket{n_0+1}_j\ket{n_0-1}_k\Pi_{l\neq ij}
\ket{n_0}_l$.  Using these expressions it is straightforward to
compute the expectation value of the density correction
\begin{equation}
\av{\tilde{A}_2} = n_0(n_0+1) \sum_{\langle j(i) \rangle}
\left(\frac{1}{(E_0-E_{ij})^2} - \frac{1}{(E_0-E_{ji})^2} \right) ,
\end{equation}
where $E_0 -E_{ij} = -U + \Delta_j -\Delta_i$. It is now
straightforward to calculate the disorder averages and the overlap
is given by
\begin{equation}
q = 64(n_0(n_0+1))^2p(1-p)z(z+1)\frac{J^4U^2\Delta^2}{(U^2 - 4\Delta^2)^4} ,
\label{eq:MI_result}
\end{equation}
where $z$ is the number of nearest neighbours.

Next we turn to the Bose-glass phase and consider the leading order
(non-vanishing) correction to the overlap function which occurs here
at second order. The expression for the overlap in lowest-order
perturbation theory is then given by
\begin{equation}
q = q_0 + J^2 \left([\av{n_i}\av{\tilde{A}_2}]_{av} -
[\av{n_i}]_{av}[\av{\tilde{A}_2}]_{av} \right),
\end{equation}
where $q_0 = p(1-p)$ (the overlap at zero hopping), $\tilde{A}_2$
and $S_1$ are as given above for the MI phase but with
\begin{equation}
\ket{\Omega}
= \prod_l \left[d_l \ket{n_0+1}_l + (1-d_l)\ket{n_0}_l \right]
\end{equation}
and
\begin{equation}
\begin{split}
\ket{\phi_{jk}} = \left[d_j\ket{n_0+2}_j + (1-d_j) \ket{n_0+1}_j
\right] [d_k\ket{n_0}_k \\ \nonumber
+ (1-d_k)\ket{n_0-1}_k ] \Pi_{l\neq
jk}[d_l \ket{n_0+1}_l + (1-d_l) \ket{n_0}_l ],
\end{split}
\end{equation}
where $d_{j,k,l}$ $(1-d_{j,k,l})$ is zero (one) for $\Delta_{j,k,l} =
\Delta$ ($\Delta_{j,k,l} = -\Delta$). Again, using these expressions
it is straightforward to compute the expectation value of the density
correction and for $n_0\geq1$ we obtain
\begin{equation}
\av{\tilde{A}_2}= \sum_{\langle j(i) \rangle}
\left(\frac{\gamma_{ij}^2}{(E_0-E_{ij})^2} -
\frac{\gamma_{ji}^2}{(E_0-E_{ji})^2} \right),
\end{equation}
where
\begin{equation}
E_0-E_{ij} = -U(1-d_i + d_j) + \Delta_j -\Delta_i
\end{equation}
and
\begin{eqnarray}
\gamma_{ij} &=& (d_i \sqrt{n_0 + 2} \\ \nonumber
&&+ (1-d_i)
\sqrt{n_0+1})(d_j\sqrt{n_0+1} + (1-d_j)\sqrt{n_0}) .
\end{eqnarray}
Once again a straightforward calculation of the disorder averages
gives the following expression ($n_0\geq 1$) for the overlap:
\begin{equation} \label{eq:q_bg_result}
q = p(1-p) \left (1 -J^2\frac{(n_0+1)^2U(U-2\Delta) + \Delta^2
}{2\Delta^2(U-\Delta)^2} \right).
\end{equation}
An analogous calculation for $n_0=0$ gives
\begin{equation} \label{eq:q_bg_result_2}
q=p(1-p) \left(1 -\frac{J^2}{2\Delta^2} \right)
\end{equation}
which corresponds to the $U \rightarrow 0$ limit of~(\ref{eq:q_bg_result}).

When the effects of hopping dominate, the bosonic atoms are in
the superfluid phase (see figure~\ref{fig:phase_diagram}). Then,
the influence of weak disorder on the superfluid phase can be
studied within a generalized mean-field approach. In analogy to the
Gross-Pitaevskii formalism, we replace the bosonic operators $b_{i}$ in
the Hamiltonian~(\ref{dBHM}) by a local complex field $\psi_{i}$,
and minimize the corresponding free energy functional $H(\psi_i)$.
The effect of disorder is to produce fluctuations in the local
density. Hence we make the ansatz $\psi_i = e^{i\phi}\sqrt{n_0 +
\delta n_i}$ where $n_0$ is the homogeneous density in the absence
of disorder and $\delta n_{i} \propto \Delta_{i}$ is the disorder
induced density fluctuation. Introducing the notation $\delta n=
[\delta n_{i} ]_{av}$ as the disorder-averaged density fluctuation,
the particle density reduces to $[\langle n_i \rangle]_{av}=
n_{0}+ \delta n$. We substitute this ansatz into $H(\psi_i)$,
expand $\psi_i$ for small fluctuations, $\delta n_i \ll n_0$ (see
appendix~\ref{app:mean_field}) and minimize the resulting expression
in~(\ref{H_density_fluctuations}) at fixed particle number, which
is equivalent to requiring $[\delta n_i]_{av}=0 $. Note that for large
dimension and thus a large number of nearest neighbours we may replace
\begin{equation} \label{crude_mean_field}
\sum_{\langle j(i) \rangle} \delta n_j \approx z \delta n.
\end{equation}
Using this relation and self-consistently solving for the density
fluctuations $\delta n_i$ (see appendix~\ref{app:mean_field}) gives
the overlap function
\begin{equation} \label{q_sf}
  q= \left(\frac{4 n_{0} \Delta}{2 U n_{0} + z J}\right)^{2} p (1-p),
\end{equation}
where $n_{0}= (\mu + J z)/U + \frac{1}{2}$.

In summary, we find that in the limit of small hopping $J/U\ll 1$,
the overlap signals a sharp crossover between the Mott-insulator
with $q\approx 0$ and the Bose-glass with $q\approx p(1-q)$,
see figure~\ref{fig:phase_diagram}. In turn, the superfluid phase is
characterized by off-diagonal long range order giving rise to coherence
peaks in a time of flight picture. Consequently, the overlap and
the coherence peaks in time of flight allow for a clear identification
of the phases in the disordered Bose-Hubbard model. However, we would
like to point out that the overlap function behaves smoothly across
the phase transition and consequently is not suitable for the precise
determination of the exact location of
the phase transition.

\subsection{Numerical results}
\label{sect:numerical_particle_overlap}

We have simulated the disordered Bose-Hubbard model~(\ref{dBHM})
in 1D using a number-conserving algorithm based on time-dependent
matrix product states~\cite{vidal:03,vidal:04}. In the matrix product
state representations, we reduce the Hilbert space by retaining only
the $\chi$ most significant components in a Schmidt-decomposition of
the state at each possible bipartite splitting of the system. States in
1D can be efficiently represented in this way, with relatively small
$\chi$ giving essentially unity overlap between the represented state
and the actual state. The simulations both serve as a check for the
perturbation theory, and allow computation of the overlap function
beyond the perturbation limit from the microscopic model.

Because we consider the regime $2\Delta<U$, the system remains
incompressible and does not enter the disordered glass phase.
Therefore we do not expect replica-symmetry breaking to occur. For
our simulations, the presence of replica-symmetry is equivalent to
convergence to a unique ground-state for any given disorder
independent of initial conditions. The overlap function can then be
computed as
\begin{equation}\label{numq}q=\frac{1}{L}\sum_i\langle
n_i\rangle^2-\overline{n}^2
\end{equation}
where $\overline{n}$ is the density of the system. Computation of
the overlap function can be simplified in this way as
$L^{-1}\sum_i \langle n_i\rangle$ self-averages to
$[\langle n_i\rangle]_{av}$ for sufficiently large systems, which
in turn self-averages to $\overline{n}$.

\begin{figure}[h!]
\includegraphics[width=8cm]{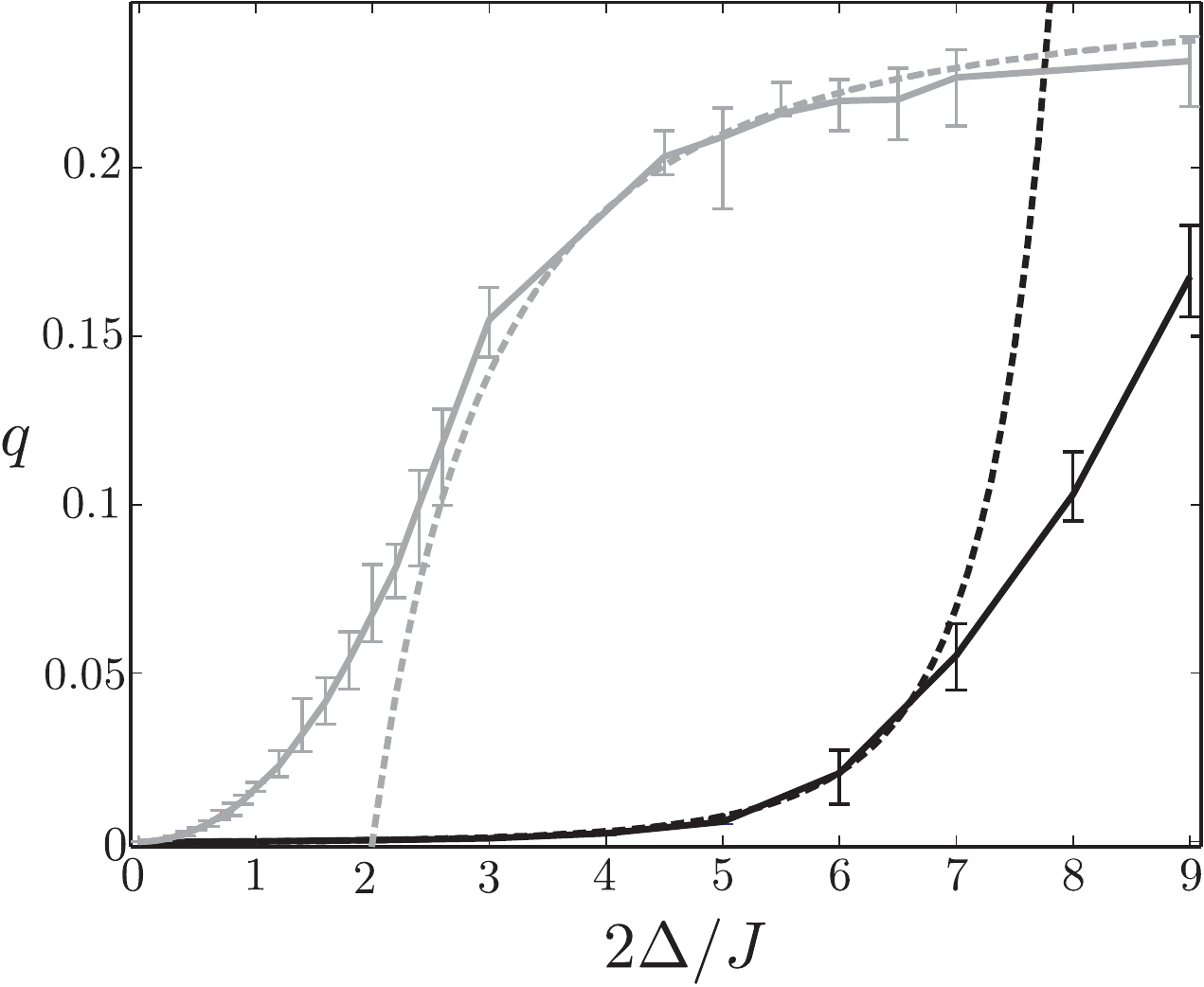}
\caption{Overlap function $q$ as a function of impurity strength
$\Delta$ in the Mott-insulating phase at filling factor $n_0=1$
(black lines) and in the Bose-glass phase for $n_0=0$ (grey lines).
In both cases, the impurity probability is $p=0.5$ and $U/J=10$,
for $L=64$. Simulation results (solid lines) are shown versus
results from perturbation theory (dashed lines). Much of the
variation in the simulated overlap function is due to the fact
that we allow the number of impurities to fluctuate around the
value given by $Lp$ in each particular disorder realization. Note
that the perturbation result in the Mott-insulator diverges as
$\Delta$ approaches $U$ due to energy degeneracy [c.f.
~(\ref{eq:MI_result})]. In the Bose-glass phase, second-order
perturbation theory yields a divergence as $\Delta$ approaches
$0$, again due to energy degeneracy [c.f.~(\ref{eq:q_bg_result_2})].}
\label{fig:combinedqfig}
\end{figure}

The ground states used in~(\ref{numq}) are computed by imaginary-time
evolution, c.f.~\cite{vidal:04}, of the 1D Bose-Hubbard-model
given by~(\ref{dBHM}) with a randomly-chosen bimodal distribution,
and different impurity strengths $2\Delta$.  The lattice size
is $L=64$, representative of current experimental capabilities,
and we compute the overlap as a function of $\Delta$ for several
different values of $U/J$ and the fraction of impurity-occupied
sites, $p$. In figure~\ref{fig:combinedqfig} we display results for
$U/J=10$ and $p=0.5$. We have performed convergence tests in the time step,
the duration of imaginary-time evolution, the number of disorder
realizations, and the number of retained states $\chi$. We find that
the overlap function converged even for surprisingly small values
of $\chi\sim20$ states in the large $U/J$ limit (Mott-insulator or
Bose-glass regime). Regarding the number of disorder realizations for
each value of $\Delta$, we find that whilst $10$ random realizations
of bimodal disorder are not sufficient to narrow the error down,
$20$ realizations suffice. Much of the observed fluctuations
in the value of the overlap function for a given $\Delta$ stem from
the fact that we allow the number of impurities in each realization
to fluctuate around the mean $Lp$. Conversely, in our simulation data
we observe different disorder realizations with the same number of
impurities having overlaps closer to each other than the error bars
is figure~\ref{fig:combinedqfig} would suggest.

As shown in figure~\ref{fig:combinedqfig}, we generally found very good
agreement between numerical simulations and perturbation theory in
its region of validity. For the Mott-insulator regime (commensurate
filling factor) this is the regime where $2\Delta\ll U$. We observe
the breakdown of our second-order perturbation theory as $\Delta$
approaches $U/2 = 5J$, which is caused by energy-degeneracy of adjacent
impurity- and non-impurity occupied sites when $\Delta=U/2$. In the
Bose-glass regime, breakdown of perturbation theory sets in for low
$\Delta$, as degeneracy occurs there for $\Delta=0$.

\section{Conclusion}

We have shown how the Bose-glass phase can be identified by measuring
the overlap function characterizing the correlations between disorder
replicas, i.e., systems having identical disorder landscape. We have
described a procedure to create disorder replicas using cold atomic
gases in an optical lattice, focusing on the case of disorder induced
by a second atomic species. A scheme to measure the overlap has been
presented, which involves the characterization of the occupation
numbers in both the individual replicas and the joint system, where
both replicas have been combined. Specifically, we have shown that
after combining the replicas together, particles can be distributed
across multiple motional states and we explained in detail how to
determine the occupation number distribution in this situation. Using
perturbation theory we have calculated the overlap for weak hopping and
have shown the different behaviours exhibited in the Mott-insulator
and Bose glass phases; these results are in good agreement with the
results from a one-dimensional numerical simulation. In the opposite
limit of very large hopping we also obtain analytical results for
the overlap within a mean-field-theory treatment.

Applying our proposed measurement scheme for the overlap to other
types of disordered models \cite{giamarchi:01,parisi:83} would provide
interesting insights to characteristic properties of disorder-induced
quantum phases, such as the possible appearance of broken ergodicity and different
degenerate ground states. Of particular interest would be the quantum spin
glass where the overlap corresponds to the Edwards-Anderson order
parameter which has been conjectured to exhibit replica symmetry
breaking \cite{parisi:83}, i.e., the results for the overlap depend
on the particular replica. Our method could also give novel insights into physics of systems that
possess multiple metastable states, such as ultracold dipolar gases \cite{menotti:07} or quantum
emulsions \cite{roscilde:07}; in particular statistics of overlaps of such metastable
states could be measured. In addition to equilibrium properties
considered here, the dynamics of the overlap might also posses an
interesting behaviour, and could be  measured in a similar setup. In
one-dimension the time evolution of the overlap could be determined
numerically \cite{vidal:04}.

\begin{acknowledgements}

Work in Innsbruck was supported by the Austrian Science foundation through
SFB F15, the EU network SCALA and project I118\_N16 (EuroQUAM\_DQS).
S.M.~thanks the Department of Physics of the University of Auckland for
their hospitality. H.G.K.~acknowledges support from the Swiss
National Science Foundation under Grant No.~PP002-114713.
M.L.~acknowledges support of ESF PESC Programme ``QUDEDIS" and
Spanish MEC grants (FIS 2005-04627, Conslider  Ingenio 2010 ``QOIT").
\end{acknowledgements}

\appendix

\section{Energy shifts and couplings for three and four particles}

In the following two sections we determine
expressions for the eigenstates and eigenenergies of
~(\ref{hamiltonian_two_bands_two_species}), i.e., $\ket{m,n}_\pm$
and $E_\pm^{(mn)}$ and associated energy shifts and couplings for
all possible initial configurations of three and four particles. Note
that the case where all particles are in the lowest or highest energy
band has already been explained in section~\ref{sect:species_transfer}
and will not be discussed here further.

\subsection{Three Particles} \label{appendix_1a}

We consider the two different configurations described by the initial
states i) $\ket{2,1;0,0}$ and ii) $\ket{1,2;0,0}$.

For case i) the two eigenstates are given by
\begin{eqnarray}
\ket{2,1}_+ & = & \frac{1}{\sqrt{3}}(\ket{2,0;0,1} + \sqrt{2} \ket{1,1;1,0}), \\
\ket{2,1}_- & = & \frac{1}{\sqrt{3}}(-\sqrt{2}\ket{2,0;0,1} + \ket{1,1;1,0})
\end{eqnarray}
with eigenenergies $E_+^{(2,1)} = 2\epsilon + 1$ and $E_-^{(2,1)} =
\epsilon/2 + 1$ respectively. We see that only the state $\ket{2,1}_+$
couples to the initial state $\ket{2,1;0,0}$, with a matrix coupling
element of $\Omega_{fi} = \sqrt{3}$, whilst for the state $\ket{2,1}_-
$ we find $\Omega_{fi}=0$. The energy shifts for the two different
possible processes are given by $\Delta E/U_{00}^{aa} = 2(\epsilon -1)$
and $\Delta E/U_{00}^{aa} = \frac{\epsilon}{2} -2$.

For case ii) the two eigenstates are given by
\begin{eqnarray}
\ket{1,2}_\pm & = &
\frac{1}{\sqrt{1+(\gamma_\pm^{(1,2)})^2}}\left(\gamma_\pm^{(1,2)}\ket{0,2;1,0} +\ket{1,1;0,1}\right), \nonumber\\
\end{eqnarray}
where
\begin{eqnarray}
\gamma_\pm^{(1,2)} & = & \frac{-1 - \epsilon \pm \sqrt{33\epsilon^2+2 \epsilon+1}}{4 \sqrt{2} \epsilon},
\end{eqnarray}
and with eigenenergies
\begin{eqnarray}
E_\pm^{(1,2)} & = & \epsilon +1 \pm \frac{1}{8}\sqrt{33\epsilon^2+2 \epsilon+1} + \frac{1}{8}(\epsilon-1). \nonumber \\
\end{eqnarray}
In this case both states $\ket{1,2}_\pm$ have nonzero
coupling to the state $\ket{1,2;0,0}$ given by
\begin{eqnarray}
\Omega_{fi} &=& \eta_\pm^{(1,2)} \equiv \bra{1,2}_\pm H_{\rs RC} \ket{1,2;0,0}
\\ \nonumber
&=& \frac{1}{\sqrt{1+(\gamma_\pm^{(1,2)})^2}}
(\gamma_\pm^{(1,2)}+\sqrt{2}) .
\end{eqnarray}
The energy shifts for the two
different possible processes are given by $\Delta E/U_{00}^{aa} =
E_\pm^{(1,2)}-3$.

\subsection{Four Particles} \label{appendix_1b}

We consider the three different configurations described by the
initial states i) $\ket{3,1;0,0}$, ii) $\ket{2,2;0,0}$ and iii)
$\ket{1,3;0,0}$.

For case i) the eigenstates are given by
\begin{eqnarray}
\ket{3,1}_+ & = & \frac{1}{\sqrt{4}}(\ket{3,0;0,1} + \sqrt{3} \ket{2,1;1,0}), \\
\ket{3,1}_-& = & \frac{1}{\sqrt{4}}(-\sqrt{3}\ket{3,0;0,1} + \ket{2,1;1,0}),
\end{eqnarray}
with eigenenergies $E_+^{(3,1)} = 3(\epsilon + 1)$ and $E_-^{(3,1)} =
\epsilon + 3$, respectively. We see that only the state $\ket{3,1}_+$
couples to the initial state $\ket{3,1;0,0}$, with a matrix coupling
element of $\Omega_{fi} = \sqrt{4}$, whilst for the state $\ket{3,1}_-
$ we find $\Omega_{fi}=0$. The energy shifts for the two different
possible processes are given by $\Delta E/U_{00}^{aa} = 3(\epsilon -1)$
and $\Delta E/U_{00}^{aa} = \epsilon -3$.

For case ii) the eigenstates are given by
\begin{eqnarray}
\ket{2,2}_\pm & = &
\frac{1}{\sqrt{1+(\gamma_\pm^{(2,2)})^2}}\left(\gamma_\pm^{2,2}\ket{1,2;1,0} +
\ket{2,1;0,1}\right), \nonumber \\
\end{eqnarray}
where
\begin{eqnarray}
\gamma_\pm^{(2,2)} & = & \frac{-1 + \epsilon \pm \sqrt{65\epsilon^2-2 \epsilon
+1}}{8 \epsilon},
\end{eqnarray}
and with eigenenergies
\begin{eqnarray}
E_\pm^{(2,2)} & = & 3 + 2 \epsilon \pm \sqrt{65\epsilon^2-2\epsilon +1} -
\frac{1}{8}(1+\epsilon). \nonumber \\
\end{eqnarray}
In this case, generally both states $\ket{2,2}_\pm$ have nonzero
coupling to the state $\ket{2,2;0,0}$ given by
\begin{eqnarray}
\Omega_{fi} &=& \eta_\pm^{(2,2)} \equiv \bra{2,2}_\pm H_{\rs RC} \ket{2,2;0,0}
\\ \nonumber
&=&
\sqrt{\frac{2}{1+(\gamma_\pm^{(2,2)})^2}} (\gamma_\pm^{(2,2)}+1) .
\end{eqnarray}
The
energy shifts for the two different possible processes are given by
$\Delta E/U_{00}^{aa} = E_\pm^{(2,2)}-\frac{23}{4}$.

For case iii) the eigenstates are given by
\begin{eqnarray}
\ket{1,3}_\pm & = &
\frac{1}{\sqrt{1+(\gamma_\pm^{(1,3)})^2}}\left(\gamma_\pm^{(1,3)}\ket{0,3;1,0} +
\ket{1,2;0,1}\right), \nonumber \\
\end{eqnarray}
where
\begin{eqnarray}
\gamma_\pm^{(1,3)} & = & \frac{-1 - \epsilon \pm \sqrt{13\epsilon^2+2 \epsilon
+1}}{2 \sqrt{3} \epsilon},
\end{eqnarray}
and with eigenenergies
\begin{eqnarray}
E_\pm^{(1,3)} & = & 3 + 2 \epsilon \pm \sqrt{13\epsilon^2+ 2\epsilon+1}
-\frac{1}{4}(2+\epsilon). \nonumber \\
\end{eqnarray}
In this case, generally both states $\ket{1,3}_\pm$ have nonzero
coupling to the state $\ket{1,3;0,0}$ given by
\begin{eqnarray}
\Omega_{fi} &=&
\eta_\pm^{(1,3)} \equiv \bra{1,3}_\pm H_{\rs RC} \ket{1,3;0,0} \\\nonumber
&=&
\frac{1}{\sqrt{1+(\gamma_\pm^{(1,3)})^2}}(\gamma_\pm^{(1,3)}+\sqrt{3}) .
\end{eqnarray}
The energy shifts for the two different possible processes are given
by $\Delta E/U_{00}^{aa} = E_\pm^{(2,2)}-\frac{15}{4}$.

\section{Perturbation Theory} \label{app:pert_theory}

In this section we derive the corrections to the overlap function
due to small hopping in perturbation theory following closely the
method presented in \cite{cohen-tannoudji:92}. We start with the
zero hopping Hamiltonian $H_0$ with ground state $\ket{\Omega}$ and
eigenenergy $E_0$. We treat the hopping Hamiltonian, $JH_1$ with $H_1 =
-\sum_{\langle i,j \rangle} b_i^{\dagger} b_j$, which has eigenstates
denoted by $\ket{\phi_{ij}}$ and eigenenergies denoted by $E_{ij}$ in
perturbation theory. We proceed by determining a unitary transformation
that transforms the total Hamiltonian $H=H_0+JH_1$ into an effective
Hamiltonian $H'$ with the identical eigenspectrum but acting only
in the unperturbed Hilbert space of the state $\ket{\Omega}$. The
required unitary transformation may be written as $U=e^{-iS}$ and can
be determined consistently to any order by writing $S = JS_1 +J^2S_2 +
\cdots$. We assume that $S$ has only non-diagonal matrix elements which
connect the unperturbed ground state $\ket{\Omega}$ and the excited
states $\ket{\phi_{ij}}$. Using the unitary transformation we can
directly calculate the correction to the particle density due to the
hopping by computing $\bra{\Omega} \tilde{n}_i \ket{\Omega}$ where
$\tilde{n}_i\equiv U^\dagger n_i U$. We show in the following sections
that it is sufficient to determine $S$ to first order. Moreover, it
is straightforward to show that to first order the matrix elements
of $S$, i.e. $S_1$, are given by
\begin{equation}
\bra{\Omega} i S_1 \ket{\phi_{ij}} = \frac{\bra{\Omega} V  \ket{\phi_{ij}}}{E_0
- E_{ij}}
\end{equation}

\subsection{Mott-insulator}

In this section we derive the leading order correction to the
overlap function for small hopping in perturbation theory for the MI
phase. We begin by expanding $S$ to fourth order and calculate the
average corrected density to be
\begin{equation} \label{eq:corrected_density}
\av{\tilde{n}_i} = \av{n_i} +\sum_{j=1}^4 J^j \av{A_j},
\end{equation}
where
\begin{eqnarray}
A_1 &=& [S_1,n_i], \\
A_2 &=& [S_2,n_i] -\frac{1}{2} [S_1,[S_1,n_i]], \\
A_3 &=& [S_3,n_i] -\frac{1}{2} \left( [S_1,[S_2,n_i]] + [S_2,[S_1,n_i]] \right)
\nonumber \\
    && -\frac{1}{6}[S_1,[S_1,[S_1,n_i]]], \\
A_4 &=& [S_4,n_i] -\frac{1}{2}[S_2,[S_2,n_i]] +
\frac{1}{24}[S_1,[S_1,[S_1,[S_1,n_i]]]]  \nonumber \\
    && -\frac{1}{2} \left( [S_1,[S_3,n_i]] + [S_3,[S_1,n_i]] \right).
\end{eqnarray}
First, note that since we have chosen $S$ to have no diagonal matrix
elements $\av{[S_n,n_i]} = 0$ to all orders.

We now calculate the first part of the overlap function as given in
~(\ref{eq:q_EA}) and find
\begin{eqnarray}
[\av{\tilde{n}_i}^2]_{av} & = & [\av{n_i}^2]_{av} + J^2[\av{n_i}\av{A_2}]_{av} +
J^3 [\av{n_i}\av{A_3}]_{av} \nonumber \\ && + J^4[(\av{n_i}\av{A_4} +
\av{A_2}^2]_{av} + \mathcal{O}(J^5).
\end{eqnarray}
Similarly, for the second term of the overlap in~(\ref{eq:q_EA})
we obtain the following expression
\begin{eqnarray}
[\av{\tilde{n}_i}]_{av}^2 & = & [\av{n_i}]_{av}^2 +
J^2[\av{n_i}]_{av}[\av{A_2}]_{av} \\ \nonumber
&& + J^3 [\av{n_i}]_{av}[\av{A_3}]_{av} \\ \nonumber
&& +J^4([\av{n_i}]_{av}[\av{A_4}]_{av} \\ \nonumber
&& + [\av{A_2}]_{av}^2) +
\mathcal{O}(J^5).
\end{eqnarray}
Using $\av{n_i} = n_0$ we obtain the following expression for the overlap
\begin{eqnarray}
q & = & [\av{\tilde{n}_i}^2]_{av} - [\av{\tilde{n}_i}]_{av}^2, \nonumber \\
  & = & J^4 \left([\av{A_2}^2]_{av} - [\av{A_2}]_{av}^2 \right) +
\mathcal{O}(J^5).
\end{eqnarray}
Note that because $\av{n_i}=n_0$ at all lattice sites (independent of the
disorder) the second- and third-order contributions cancel.

To determine the required commutators $[S_1,[S_1,n_i]]$ we write $S_1$
and $n_i$ as projectors
\begin{eqnarray}
S_1 &=& -i \sum_{\av{jk}} c_{jk} \ket{\Omega}\bra{\phi_{jk}} + \textrm{H.c.}, \\
n_i &=& n_0\ket{\Omega}\bra{\Omega} + n_0\sum_{\av{jk} \neq i}
\ket{\phi_{jk}}\bra{\phi_{jk}}  \nonumber \\
    && + (n_0+1)\sum_{\langle j(i) \rangle} \ket{\phi_{ij}}\bra{\phi_{ij}}
\nonumber \\
    && + (n_0-1)\sum_{\langle j(i) \rangle} \ket{\phi_{ji}}\bra{\phi_{ji}},
\end{eqnarray}
where
\begin{equation}
c_{jk} = \frac{\bra{\Omega} V \ket{\phi_{jk}}}{E_0 -E_{jk}},
\end{equation}
\begin{equation}
\ket{\Omega} =\prod_i \ket{n_0}_i,
\end{equation}
and
\begin{equation}
\ket{\phi_{jk}} =\ket{n_0+1}_j\ket{n_0-1}_k\prod_{l\neq i,j}\ket{n_0}_l .
\end{equation}
Using these
expressions it is straightforward to compute the expectation value of the
density correction
\begin{equation}
\av{A_2} = n_0(n_0+1)\sum_{\langle j(i) \rangle} \left(\frac{1}{(E_0-E_{ij})^2}
- \frac{1}{(E_0-E_{ji})^2} \right),
\end{equation}
where $E_0 - E_{ij} = -U+\Delta_j-\Delta_i$.

\subsection{Bose-glass}

In this section we derive the leading order correction to the
overlap function for small hopping in perturbation theory for the BG
phase. Again we consider the corrected density $\av{\tilde{n}_i}$ as
given in~(\ref{eq:corrected_density}) but only up to second order
in $J$. Again we have $\av{[S_n,n_i]} = 0$ because we have chosen $S$
to have no diagonal matrix elements. Thus the overlap to second
order is given by
\begin{eqnarray}
q & = & [\av{\tilde{n}_i}^2]_{av} - [\av{\tilde{n}_i}]_{av}^2 \nonumber \\
 & = & q_0 -J^2 \left([\av{n_i}\av{A_2}]_{av}-
[\av{n_i}]_{av}[\av{A_2}]_{av}\right) + \mathcal{O}(J^3), \nonumber \\
\end{eqnarray}
where $q_0 = [\av{n_i}^2]_{av}-[\av{n_i}]_{av}^2$
is the overlap at zero hopping derived in
section~\ref{sect:analytical_particle_overlap}. Since $\av{n_i}$ is now
site dependent the second-order contribution does not vanish.

To determine the required commutators $[S_1,[S_1,n_i]]$ we again
write $S_1$ and $n_i$ as projectors
\begin{eqnarray}
S_1 &=& -i \sum_{\av{jk}} c_{jk} \ket{\Omega}\bra{\phi_{jk}} + \textrm{H.c.}, \\
n_i &=& (n_0+d_i) \ket{\Omega}\bra{\Omega} + (n_0+d_i)\sum_{\av{jk} \neq i}
\ket{\phi_{jk}}\bra{\phi_{jk}}  \nonumber \\
    && + (d_i(n_0+2)+(1-d_i)(n_0+1))\sum_{\langle j(i) \rangle}
\ket{\phi_{ij}}\bra{\phi_{ij}}  \nonumber \\
    && + (d_in_0+(1-d_i)(n_0-1))\sum_{\langle j(i) \rangle}
\ket{\phi_{ji}}\bra{\phi_{ji}},
\end{eqnarray}
where
\begin{equation}
c_{jk} = \frac{\bra{\Omega} V \ket{\phi_{jk}}}{E_0
- E_{jk}} ,
\end{equation}
with
\begin{equation}
\ket{\Omega} = \prod_l [d_l\ket{n_0}_l +(1-d_l)\ket{n_0+1}_l]
\end{equation}
and
\begin{eqnarray}
\ket{\phi_{jk}} &=& [d_j\ket{n_0+2}_j +
(1-d_j) \ket{n_0+1}_j][d_k\ket{n_0}_k \nonumber  \\
&& +(1-d_k)\ket{n_0-1}_k]  \nonumber \\
&& \times \Pi_{l\neq
jk} [d_l \ket{n_0+1}_l + (1-d_l) \ket{n_0}_l]
\end{eqnarray}
where $d_{j,k,l}$ ($1-d_{j,k,l}$) is zero (one) for $\Delta_{j,k,l} = \Delta$
($\Delta_{j,k,l} = -\Delta$). Again, using these expressions it is
straightforward to compute the expectation value of the density
correction and for $n_0\geq1$ we obtain
\begin{equation}
\av{A_2} = \sum_{\langle j(i) \rangle} \left(
\frac{\gamma_{ij}^2}{(E_0-E_{ij})^2} - \frac{\gamma_{ji}^2}{(E_0-E_{ji})^2}
\right)
\end{equation}
where
\begin{eqnarray}
\gamma_{ij} &=& \left(d_i \sqrt{n_0 + 2} + (1-d_i) \sqrt{n_0+1}\right)
        \left(d_j\sqrt{n_0+1}\right.
\nonumber \\
 && + \left. (1-d_j)\sqrt{n_0}\right)
\end{eqnarray}
and $E_0 - E_{ij} = -U[d_i+(1-d_j)]+\Delta_j-\Delta_i$.

\section{Mean Field Theory}
\label{app:mean_field}

We replace the annihilation operators $b_i$ in the Hamiltonian in
~(\ref{dBHM}) by the mean field $\psi_i = e^{i\phi}\sqrt{n_0
+ \delta n_i}$ and expand the fluctuations $\delta n_i \ll n_0$
up to second order to obtain
\begin{eqnarray} \label{H_density_fluctuations}
H (\delta n_i)  & = & - \frac{J}{4n_0} \sum_{\av{ij}} \delta n_i \delta
n_j - \frac{1}{2}(\delta n_j^2 +\delta n_i^2) \nonumber \\
&& + \frac{U}{2} \sum_i \delta n_i^2 + (2n_0-1) \delta n_i \nonumber \\
&& - \sum_i (\mu+zJ-\Delta_i)\delta n_i + \mathcal{O}(\delta n_i^3).
\end{eqnarray}
Next, we minimize~(\ref{H_density_fluctuations}) at fixed particle
number which is equivalent to requiring $[\delta n_i]_{av}=0 $.
Using~(\ref{crude_mean_field}) and consistently solving gives
\begin{equation} \label{density_fluctuation}
\delta n_i = \frac{\Delta (2p-1) - \Delta}{U+\frac{zJ}{2n_0}}.
\end{equation}
Finally, then the overlap can be expressed in terms of these
fluctuations as follows
\begin{eqnarray} \label{q_density_fluctuation}
q & = & [\av{n_i}^2]_{av} - [\av{n_i}]_{av}^2 \nonumber\\
  &=  & [\delta n_i^2]_{av} - [\delta n_i]_{av}^2.
\end{eqnarray}

Substituting~(\ref{density_fluctuation}) into
~(\ref{q_density_fluctuation}) and performing the disorder
average for a bimodal distribution gives the result in~(\ref{q_sf})
for the overlap function $q$.
\bibliographystyle{apsrevtitle}
\bibliography{refs,comments}

\begin{thebibliography}{70}
\bibitem[{com({\natexlab{a}})}]{comment:overlap}
\bibinfo{note}{Note that our overlap function is defined as the averaged
  squared deviation from the mean. That is why, while the overlap of the two
  Mott replicas is indeed perfect, our overlap function, defined via
  deviations, is zero.}
\bibitem[{com({\natexlab{c}})}]{comment:stats}
\bibinfo{note}{For small systems it may be necessary to improve the statistics
  by a repetition of the measurement.}
\bibitem[{com({\natexlab{d}})}]{comment:occupation}
\bibinfo{note}{In case the occupation number distribution in each replica is
  not equal, each should be individually addressed and measured, e.g., using a
  magnetic field gradient during the measurement.}
\bibitem[{com({\natexlab{e}})}]{comment:table}
\bibinfo{note}{These points can easily be determined analytically by equating
  energy shifts for different total particle numbers (as given in Table
  \ref{table:species_transfer}) and solving for $\epsilon$.}
\end{thebibliography}

\end{document}